\documentclass[useAMS,usenatbib]{mn2e}

\title[A search for ultracool dwarfs in moving groups]{A search for southern ultracool dwarfs in young moving groups}
\author[J. R. A. Clarke et al]{J. R. A. Clarke$^1$,\thanks{e-mail: j.r.a.clarke@herts.ac.uk} D. J. Pinfield$^1$, M. C. G$\rm{\acute{a}}$lvez-Ortiz$^1$, J. S. Jenkins$^2$, 
\newauthor  B. Burningham$^1$, N. R. Deacon$^{3,4}$, H. R. A. Jones$^1$, R. S. Pokorny$^{5}$, J. R. Barnes$^1$ 
\newauthor and A. C. Day-Jones$^2$ \\
$^1$Centre For Astrophysics Research, University of Hertfordshire, College Lane, Hatfield, Hertfordshire, AL10 9AB, UK\\
$^2$Department of Astronomy, Universidad de Chile, Casilla Postal 36D, Santiago, Chile\\
$^3$Department of Astrophysics, Faculty of Science, Radboud University Nijmegen, PO Box 9010, 6500 GL Nijmegen, The Netherlands \\
$^4$Institute for Astronomy, University of Hawaii, 2680 Woodlawn Drive, Honolulu, HI 96822\\
$^5$Yunnan Observatory, P. O. Box 110, CAS, 650011 Kunming,  P. R China}

\usepackage{epsfig}
\usepackage{lscape} 
\usepackage{epsfig}
\usepackage{rotating}
\usepackage{setspace}
\begin{document}

\maketitle

\label{firstpage}

\begin{abstract}

We associate 132 low-mass ultracool dwarfs in the southern hemisphere as candidate members of five moving groups using photometric and astrometric selection techniques.  Of these objects, we present high resolution spectroscopy for seven candidates and combine these with previous measurements from the literature to determine spectral types and radial velocities.  We thus constrain distance and space motion spectroscopically, allowing the kinematic membership of the moving groups to be assessed.  Possible membership of moving groups has allowed ages and metallicities to be constrained for these objects and evolutionary models have been used to estimate their mass.  We estimate that up to $\sim$75 of our candidate moving group members should be genuine, and discuss future work that will confirm and exploit this major new sample.

\end{abstract}


\section{Introduction}

Moving group (MG) populations are distinguishable from the field by their astrometric properties. A MG remains kinematically distinct within the general field population at ages $\sim$1 Gyr, before being dispersed by disk heating mechanisms (e.g. De Simone, Xiaoan \& Tremaine 2004). MG populations are thought to originate in the same formation environment as open clusters. As the progenitor gas is cleared by OB star winds and the natal cluster expands, stars with sufficiently high velocities become unbound and form a young, coeval MG, possibly leaving behind a bound open cluster (Kroupa Aarseth \& Hurley 2001), before dispersal after up to $\sim$1 Gyr.  MGs therefore consist of young populations with characteristic space motions, and membership of such groups can be used to constrain the age and composition of members.

A great deal of early work studying MGs was done by Eggen (e.g. Eggen, 1958). Although many of the older groups identified in Eggen's work have since been shown to be spurious (e.g. Taylor 2000), with their velocity structure having other causes (e.g. the effects of stochastic spiral waves; Nordstr\"{o}m et al. 2004), a variety of studies have shown that in general, the younger MGs are consistent with coeval populations of uniform metallicity (e.g. Soderblom \& Mayor 1993; Feltzing \& Holmberg 2000; Montes et al. 2001; King et al. 2003; King \& Schuler 2005; De Silva et al. 2006).

Ultracool dwarfs include objects with spectral type $\ge$M7. For the late M dwarfs (Jones \& Tsuji 1997), atmospheric dust formation has important effects on opacity and general properties. The later T dwarf spectra are dominated by strong methane and water absorption (Burgasser et~al. 2002). Ultracool dwarf masses are dependent on age, due to their significant luminosity and $T_{\rm eff}$ evolution, and they include both stellar and substellar (mass $<$0.075 M$_{\odot}$; Baraffe et~al. 1998) objects.

The identification of ultracool dwarfs in MGs facilitates a variety of interesting areas of astronomical study.

\begin{enumerate}

  \item Ultracool MG members will provide excellent test-beds (with known age and metallicity) with which to improve our theoretical understanding of low temperature and low gravity dusty-atmospheres. Such an understanding is a major stepping stone to properly characterising local low-mass populations (via spectroscopic fitting), and thus measuring the low-mass/substellar initial mass function and formation history (see Pinfield et al. 2006).

  \item Nearby intrinsically faint young stars facilitate particularly sensitive searches for binary companions over a large range of separation and mass-ratios. Studies of the properties of low-mass binary systems (Guenther et al. 2001; Luhman 2004; Delgado-Donate et al. 2004; Feigelson et al. 2006) are vital to our understanding of dynamical processes that could be important in their formation (e. g. Whitworth \& Stamatellos 2006).

  \item Close low-mass binary systems in MGs could yield dynamical mass measurements, which could be used to test evolutionary models of low-mass stars and substellar objects (Kenworthy et al. 2001; Ribas, 2003; Luhman, McLeod \& Goldenson 2005; Pinfield, Jones \& Steele 2005), providing a test for our understanding of their interior physics.

  \item The youth and potential proximity of MG members could also allow one to search for planetary companions (e. g. Chauvin et al. 2004; see also Zuckerman et al. 2001; Nakajima et al. 2005) via adaptive optics (AO) techniques (e.g. Biller et al. 2006). Around young systems, giant planets will be brighter, and can be studied at relatively early evolutionary time (see Zuckerman \& Song 2004). Furthermore, the intrinsic faintness of ultracool dwarfs reduce the contrast ratio for AO imaging at close separation (caused by the point-spread function of the primary). In essence, one can probe for lower mass planets around fainter dwarf hosts.

\end{enumerate}

Recent work (e.g. Montes et al. 2001; Lopez-Santiago et al, 2006) has produced some major new samples of nearby stars in MGs, in the early--F to M spectral type range. However, the number of ultracool dwarfs associated with MGs is still limited to a hand full of objects (Ruiz, Takamiya \& Roth 1991; Ribas, 2003; Bannister \& Jameson 2005; Barrado Y Navacu\'{e}s, 2006; Kirkpatrick et al. 2006).

In this paper we present the first stages of our new search for ultracool dwarfs in MGs in the solar neighbourhood. In the first part of the paper we focus on the creation of our red object catalogue from which MG candidates will be selected (Section 2), and describe the astrometric and photometric selection techniques with which we identify our MG candidates (Section 3). We then discuss new radial velocities and space motions in the context of kinematic membership in Section 4 and provide a first-pass assessment of the credence of our MG candidates, describing how spectroscopic methods can further constrain the membership of the objects, providing examples.  Section 5 gives a summary and discusses our planned future work with our new candidates.

\section{High Proper Motion Red Object Catalogue}

\subsection{Proper Motion Catalogues}

Large area proper motion surveys generally search for proper motions $>$0.4--0.5 arcsec/yr (Lepine, Shara \& Rich 2002; Scholz et al. 2002; Deacon, Hambly \& Cooke 2005; Lepine 2005; Subsavage et al. 2005), and are biased towards finding nearby and high velocity objects. A survey over a greater range of proper motions and with good sensitivity to red objects would be more suitable for our purposes here.

The Liverpool-Edinburgh High Proper Motion survey (LEHPM) of Pokorny, Jones \& Hambly (2003) and Pokorny et al. (2004) covers proper motions $>$0.18 arcsec/yr, and was produced using SuperCOSMOS digitised R-band ESO and UK Schmidt Plates. However, the frequency of spurious faint proper motion detections (due to various plate defects such as grains, bubbles in the emulsion, dust specks and scratches) requires a magnitude limit that is significantly brighter than the Schmidt plate limits. The LEHPM is limited to R$\sim$19.5, which is quite restrictive when searching for red faint objects.

The Two Micron All Sky Survey (2MASS) can be brought to bear directly on these issues, both by providing an extra epoch to aid the removal of spurious proper motion objects, and also to provide optical-near infrared colours with which to select red objects. 

We used two sources to compile our high proper motion red object catalogue. The first of these represents an extension of the LEHPM catalogue (which we refer to as ELEHPM), and is focused on obtaining a good accuracy for the proper motions using the large baselines available between the R-band Schmidt plate epochs. However, an expanded proper motion range (to $>$0.1 arcsec/yr) and a much deeper magnitude limit of R=21 is used. The proper motions were derived using the same methods of Pokorny, Jones \& Hambly (2003), and as expected, the initial optical list contained a very large number of spurious objects. These were dealt with to a large extent using an additional 2MASS epoch (see next Section). To compliment this R-band limited sample, we also selected from the Southern Infrared Proper Motion Survey (SIPS) catalogue (Deacon Hambly \& Cooke 2005; Deacon \& Hambly 2007). This catalogue suffers in part from short time base lines (giving less accurate proper motions), but is I-band limited (I=18) and thus photometrically more sensitive to very red objects than our R-band limited sample. We selected objects from both catalogues where $\mu/\Delta\sigma$$_{\rm{\mu}}>4$ ($\mu$ is proper motion and $\sigma_{\rm{\mu}}$ is the proper motion one sigma uncertainty) to ensure our desired level of proper motion accuracy.

There is some overlap between the two samples, but it is not as significant as might be thought.  Sixteen candidates appear in both the ELEHPM and SIPS but with proper motions that are in agreement (to 1$\sigma$).  The SIPS objects have somewhat incomplete coverage due to proper motion uncertainties above our limit resulting from the shorter available SIPS baselines. Where the proper motions are accurate enough however, the SIPS sensitivity probes more deeply for very red objects like L dwarfs.

We have also assessed the work of Faherty et al. (2009) who found that six of the SIPS objects have discrepancies with other catalogues.  We instead use these proper motions when compiling our red object catalogue.

\subsection{2MASS Cross Match and Colour Cuts}

We visually checked the SIPS catalogue and found it to be free of spurious objects, however our R-band limited sample was initially heavily contaminated.  Many of these spurious sources could be removed by requiring an additional epoch detection in the 2MASS database. 

The epoch span of the optical and 2MASS measurements cover baselines of up to 25 years.  So, to cross-match the sources the optical objects' proper motion was taken into account to derive the expected 2MASS coordinates. For each expected 2MASS position, we defined a search radius to account for uncertainties in this position. The size of the search radius depended on the accuracy of the SuperCOSMOS/2MASS coordinates, the epoch difference, and the uncertainty associated with the proper motion, with typical search radii of $\sim$5 arcseconds. 

Colour cuts requiring J-K$_s\geq$1.0 and R-K$_s\geq$5.0 were also employed using the structured query language (SQL) interface of the 2MASS GATOR, to ensure that objects with a spectral type predominantly later than M6 were returned (e.g. Kirkpatrick et al. 1999). We also required J$\leq$16 (to avoid matches with very low SNR signals within 2MASS) and that the database flag parameterising source contamination and confusion ($cc\_flg$) was $000$, which together indicate a quality detection with signal-to-noise $>10$ (e.g. Pokorny 2004). Our cross matching returned a total of 381 objects from our R-band limited sample.  These were visually inspected to identify any 2MASS counterparts that were clearly associated with different optical sources, as well as some very low signal-to-noise optical counterparts that appeared to be noise.  Of the 381 cross-matched objects, 258 passed the visual inspection.  The same colour, magnitude and quality constraints were applied to the SIPS catalogue objects, adding an additional 559 objects to our red catalogue.  There are thus 817 objects in our final high proper motion red object catalogue.  

Figure 1 shows an R versus R-K$_s$ colour magnitude diagram for our catalogue. We over-plot the R-band limit of this new catalogue (dashed line), as well as that of the LEHPM (dotted line) for comparison.    

\subsection{Cross Matching With DENIS}

Although the Deep Near Infrared Survey of the Southern Sky (DENIS) does not cover as much sky, it offers I-band photometry that is more accurate than SuperCOSMOS.  Accuracies of $<$0.1 mag are typical of DENIS I-band photometry (Epchtein et al. 1997), SuperCOSMOS has absolute photometric uncertainties of $\sim$0.3 mag.  We thus cross-matched our red object catalogues with the DENIS database using the VizieR interface tool at CDS, selecting objects within a search radius of 10 arcseconds (to allow for high proper motion) and a DENIS J-band magnitude within 0.3 magnitudes of the 2MASS detection to minimise spurious matches. In total, 482 red object catalogue sources had a DENIS counterpart (out of a total of 817) in our cross-match, and we use this DENIS photometry in preference to the SuperCOSMOS I-band measurements where it is available.

\begin{figure}
\includegraphics[angle=90,width=8.4cm]{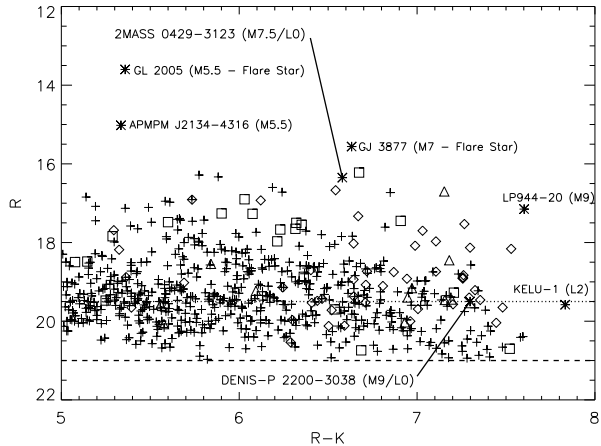}
\caption{R versus R-K colour-magnitude diagram for the new high proper motion red object catalogue. Catalogue objects are plotted as small crosses. Several previously studied interesting objects within the survey volume are highlighted as asterisks. We plot objects with known spectral types M5 - M6.5, M7 - M8.5 and M9+ as squares, diamonds and triangles respectively. The dashed and dotted lines show the R-band magnitude limits for the new catalogue and the LEHPM respectively.}
\label{roccmd}
\end{figure}

\section{Selecting The Moving Group Candidates}

\subsection{The Moving Groups}

We chose to consider the five MGs assessed in Montes et al. (2001), namely the Pleiades, Castor, Hyades, Sirius (also known as Ursa Major) and IC2391 MGs. These are relatively well studied, with previous work producing lists of kinematic members (Eggen 1983a,b; Eggen 1991; Montes et al. 2001; Lopez-Santiago et al. 2006).

Table 1 summarises the physical and astrometric properties of these MGs. Properties are generally taken from Montes et al. (2001) which represents a recent and broad assessment of these groups.  However, we have included updates where available, and also present metallicity estimates for the MGs. Space velocities UVW are given, where a positive U is towards the galactic centre, a positive V is in the direction of galactic rotation, and a positive W is in the direction of the galactic north pole. It should be noted that the Pleiades MG age has a significantly larger range than the other MGs. This is due to this MG actually being made up of a number of separate components with a range of ages (see Asiain, Figueras \& Torra 2000; Makarov 2003; Lopez-Santiago et al. 2006), although we treat the MG as a single component in this work.

\begin{table}
\label{MGs}
\begin{minipage}{87mm}
\caption{The five MGs and their properties. Properties are from Montes et al. (2001) unless specified.}
\begin{tabular}{lccccc}
\hline
MG  & Age & Metallicity & U & V & W \\ 
 & (Myr) & [Fe/H] &  &(kms$^{-1}$) & \\

\hline

Pleiades & 20-150 & -0.034$\pm$0.024$^1$ & -11.6 & -21.0 & -11.4  \\
Hyades   & 600 & 0.13$\pm$0.01$^2$ & -39.7 & -17.7 & -2.4         \\
Sirius   & 300 &  -0.09$\pm$0.04$^1$ & 14.9 & 1.0 & -10.7         \\
IC2391   & 35-55 & -0.03$\pm$0.07$^3$ & -20.6 & -15.7 & -9.1      \\
Castor   & 320$^4$ &  0.00$\pm$0.04$^5$ & -10.7 & -8.0 & -9.7     \\
\hline

\end{tabular}
\\
$^1$ Boesgaard and Friel 1990, $^2$ Paulson, Sneden and Cochran 2003, $^3$ Randich et al. 2001, $^4$ Ribas 2003, $^5$ Paulson and Yelda 2006\\
\end{minipage}
\end{table}

\subsection{Astrometric Selection}

Objects that share the same space motion must all move towards some convergent point (CP) on the sky (the vanishing point of the space motion vector).  To easily identify such objects, one must convert the two usual proper motion vectors (orientated in the RA and DEC directions) into a new coordinate system, with the CP (of the MG in question) at the pole. The two new proper motion vectors will thus represent the proper motion towards the CP ($\mu_{\rm{tcp}}$), and the proper motion perpendicular to this direction ($\mu_{\rm{pcp}}$). We used transformation equations based on the derivations of Reid (1992) to achieve this.  

\begin{figure}
\includegraphics[angle=90,width=8.4cm]{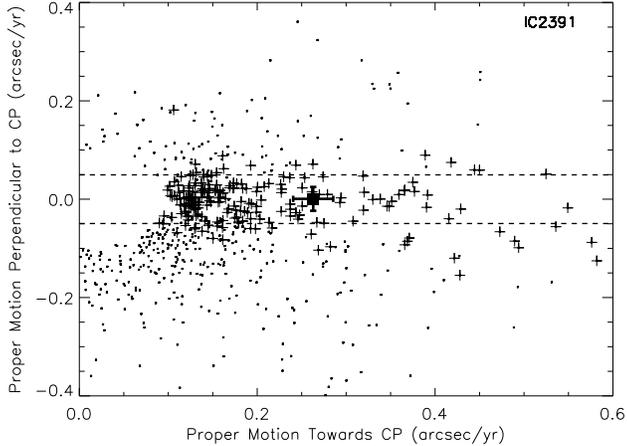}
\caption{Plot showing the converted proper motion values $\mu_{\rm{tcp}}$ and $\mu_{\rm{pcp}}$ for our catalogue of objects, when considering the IC2391 MG. We have allowed for measurement uncertainty in the proper motions, and an intrinsic scatter of $\pm$5kms$^{-1}$ in the space motion of the MG (see text).  Objects that pass the astrometric selection criteria are highlighted as plus signs.  Proper motion uncertainty and the expected intrinsic proper motion scatter are also indicated for DENIS-P J0021.0-4244 (filled square).}
\label{ic2391ast}
\end{figure}

As an example, Figure \ref{ic2391ast} shows the converted proper motions for our ultracool dwarf catalogue objects when considering the IC2391 MG.

Simplistically, one would expect MG members to all have extremely small values of $\mu_{\rm{pcp}}$, with their motion being essentially towards the CP. However, there are several factors that will produce a scatter in $\mu_{\rm{pcp}}$ about zero. Firstly, measurement uncertainty associated with the proper motions will contribute to a scatter. However, one also expects an intrinsic scatter in the space motions themselves, since a MG does not have an exact space motion. Both the (relatively slow) expansion velocity of the MG (a few kms$^{-1}$; Kroupa, Aarseth \& Hurley 2001), and any gravitational influence of disk star interactions (disk heating) will contribute to an intrinsic scatter.

To take account of these effects, we allowed for a scatter of $\pm$5kms$^{-1}$, which we converted into a proper motion velocity scatter using an astrometrically estimated distance. This distance was estimated assuming MG membership, when the total space motion of the group and the value of $\lambda_{\rm{s}}$ (angular distance between an object and the CP) for an object give the expected tangential velocity. Comparison of this velocity with the candidates total measured proper motion then yields an estimate of its distance assuming that the object is a MG member. This distance $(\it{d})$ is then, 

\begin{equation}
\label{astrodist}
d (\rm{pc}) = \frac{V_{\rm{S}} \rm{sin}(\lambda_{\rm{s}})}{k\mu_{tcp}}.
\end{equation}

Here $V_{\rm{S}}$ is the total space motion of the MG in question (in kms$^{-1}$) and $\it{k}$=4.74, equivalent to  1 AU/yr in units of kms$^{-1}$. An object was astrometrically selected as a possible MG member if its $\mu_{\rm{pcp}}$ was less than the estimated scatter or its 1-$\sigma$ proper motion uncertainties overlapped with this region (corresponding to $\pm$5kms$^{-1}$ at the astrometrically estimated distance). In the case of IC2391, such sources are highlighted in Figure \ref{ic2391ast} as plus signs. As a specific example, Figure \ref{ic2391ast} also shows the proper motion uncertainties and expected intrinsic proper motion scatter estimated for one example object, DENIS-P J0021.0-4244 (plotted as a solid square), a high proper motion object that we identify as a possible kinematic IC2391 MG member. 

\subsection{Photometric Selection}

\label{photo}

\begin{figure}
\includegraphics[angle=90,width=8.4cm]{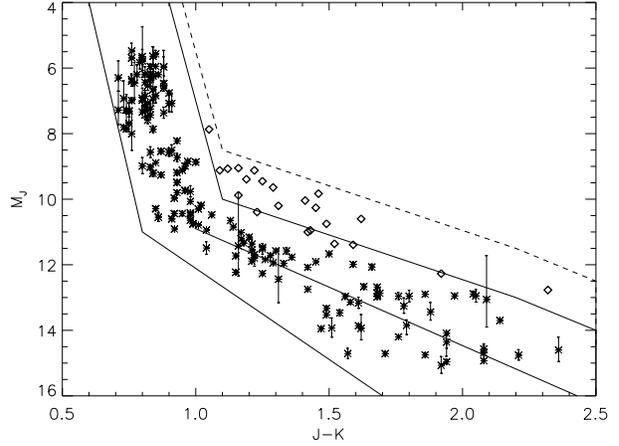}
\caption{M$_J$ versus J-K CMD of known late M and early L dwarfs with measured parallax distances plotted as asterisks.  A solid best fit line for these objects is over-plotted. The region either side of this was defined as our ultracool dwarf sequence (bounded by solid lines). Note there are two alternative upper boundaries, one brighter by 1.5 magnitudes to account for the potential youth of many of our selected objects (see text).  Example young Upper Scorpius members are plotted as diamonds.}
\label{knownobj}
\end{figure}

By applying the astrometric test described in the previous section, we were able to identify catalogue objects whose proper motion was orientated towards the convergent points of the MGs considered here (with some allowance made for both measurement uncertainty and intrinsic scatter in the astrometry). However, during this process, we also estimated a set of distances (see equation \ref{astrodist}) for the objects based on the assumption that they are members of our MGs (which allowed us to assume their space motion, and compare this to their measured astrometry). We can use these estimated distances to plot objects on a series of colour-absolute magnitude diagrams (CMDs), and thus investigate if their location on these CMDs is consistent with expectations. Objects whose location on a CMD is not consistent can then be ruled out as possible members of that particular MG.

To assess if an object's CMD location is consistent we defined a dwarf sequence on two CMDs using samples of canonical objects covering the full colour (and spectral type) range of our catalogue objects, but with distances from parallax measurements. For this purpose we used M and L dwarf samples from Leggett (1992), Dahn et al. (2002), Knapp et al. (2004) and Vrba et al. (2004).  Figure \ref{knownobj} shows the M$_{\rm{J}}$, J-K CMD for this parallax sample, where M$_{\rm{J}}$ uncertainties result primarily from the parallax uncertainties.  J and K-band photometry is available for all our red catalogue objects, and this near infrared CMD could thus be used to photometrically test all astrometric candidates.  I-band photometry was available for our catalogue objects so we also used a selection criteria employing an M$_{\rm{J}}$ versus I-J CMD (not shown here).  Thus, objects were required to pass the photometric tests for both CMDs.

\begin{figure}
\includegraphics[angle=90,width=8.4cm]{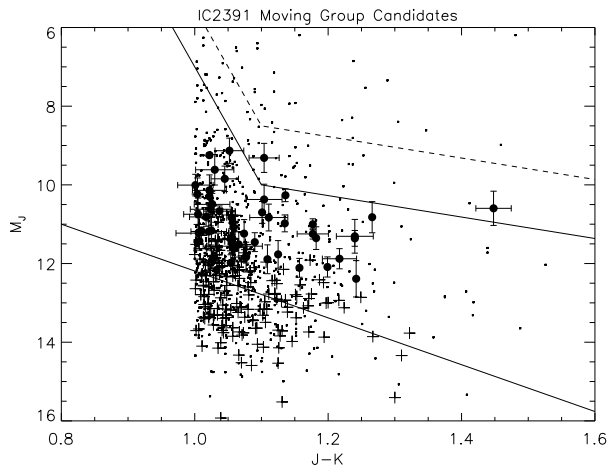}
\caption{An example M$_{\rm{J}}$, J-K CMD selection for the IC2391 MG.  The expected location of the dwarf sequence is enclosed by solid lines with the alternative upper boundary (for young sources) shown as a dashed line.  Catalogue objects 
are shown as points. Objects passing the astrometric selection (see Section 3.2) are shown as plus signs. Objects also passing our photometric selection are shown as filled circles.}
\label{ic2391phot}
\end{figure}

In order to define CMD regions that represent the dwarf sequences in this range, we initially considered boundaries that fully (and conservatively) enclose the parallax sample, allowing for the inherent scatter. Much of this scatter presumably results from unresolved binarity and some spread in metallicity causing a scatter in both M$_{\rm{J}}$ and J-K colour (e.g. fig. 7 of Leggett, Allard \& Hauschildt (1998)). However, an age spread is also a potential cause of scatter in this CMD, where some M$_{\rm{J}}$ values could be brighter due to younger objects not being fully contracted.

The parallax sample consists of bright field objects, with likely ages in the range $\sim$1--5Gyr in general. We would therefore not expect significant scatter from a large number of young objects. However, all of our MGs are younger than this (e.g. IC2391 and the Pleiades MGs in particular).  Theoretical isochrones (e.g. Baraffe et al. 1998) suggest that $\sim$20--50Myr populations could be $\sim$1--1.5 magnitudes brighter than more typical ($\sim$1--5Gyr) field star samples, and we have therefore defined alternative upper boundaries in the two CMDs, to include any potentially brighter younger objects.  This is also seen observationally; as an example we have plotted 21 BDs in the Upper Scorpius association (from Lodieu et al. 2008) as diamonds on Figure \ref{knownobj}.  The association's age of 5Myr is similar to the younger component of the Pleiades so we would expect a similar shift in brightness.  In our final candidate selections, we use the fainter upper boundaries when selecting Hyades MG candidates, and the brighter upper boundary when selecting candidate members of the other four MGs.

Figure \ref{ic2391phot} shows an example of our photometric selection for the IC2391 MG.  Catalogue objects are shown as small dots.  Objects passing the astrometric selection are highlighted as plus signs, and those that also pass our photometric selection are shown as filled circles.  Note that in Figure \ref{ic2391phot} a number of objects that pass the astrometric tests lie in the selection region of the CMD.  Although these appear within the M$_{\rm{J}}$ versus J-K selection region, they fail to be photometrically selected in the M$_{\rm{J}}$ versus I-J CMD. 

\subsection{The Moving group Candidates}

\label{mgcandidates}
By considering each MG in turn we have identified 132 objects that are candidate members of one or more of the five MGs from Table 1; 47, 89, 11, 54 and 32 objects have candidature of the Pleiades, Hyades, Sirius, IC2391 and Castor MGs respectively. Table \ref{mgcandtable} lists the candidates and gives their properties. The table also contains distance constraints for the candidates. Parallax distances are given where available. However, in the majority of cases parallaxes were not available so photometric distances (d$_{\rm{phot}}$) were estimated using an $M_{\rm{J}}$-(J-K) relation from our best fit line on the $M_{\rm{J}}$ versus J-K CMD (Figure \ref{knownobj}) with some allowance for age uncertainty (i.e. possible membership or not of one or more MGs). We describe this method further in Section \ref{spacemotions} where the same method has been used to provide distance estimates for the complete sample.

We can make an approximate estimate of the number of our MG candidates that we expect to be genuine, based on our knowledge from previous kinematic studies of the local disk population, and some assumptions about disk-age spread and structure. For a uniform disk age distribution from 0-10 Gyr, we would expect 6\% of disk objects to have ages $<$600Myr.  However, if stellar/sub-stellar birth rate is constant (e.g. Reid et al. 1999, Burgasser 2004), some bias towards younger objects would be expected due to an increased disk scale height (H$_z$) for older populations. The exact form of this H$_z$ variation is not well constrained, but H$_z$(t) could change by up to a factor of $\sim$5 (Rocha-Pinto et al. 2000; Just 2003; Deacon \& Hambly 2006; Pinfield et al. 2006) across the full disk age range. This would result in a factor $\sim$3 increase in the relative number of $<$600Myr objects. Therefore, we may expect up to $\sim$18\% of our red object catalogue to have ages in this range, some $\sim$150 objects. While not all these young objects will be in MGs, kinematic studies based on Hipparcos (Dehnen 1998) indicate that approximately 50\% of such young objects would be expected to reside in kinematic groups, and therefore we may expect $\sim$75 objects from our red object catalogue to be genuine MG members.

\begin{landscape}
\begin{table}
\begin{footnotesize}
\caption{ Properties of the 132 MG candidates.  J, H and K-band photometry is from 2MASS.  R-band photometry is from SuperCOSMOS.  I-band photometry is from DENIS where available, or from SuperCOSMOS.  Distances are parallaxes where available, or are estimated using an $M_{\rm{J}}$-(J-K) relation from our best fit line on the $M_{\rm{J}}$ versus J-K CMD with some additional allowance for age uncertainty (See Section \ref{mgcandidates}). Spectral types are from the literature where available, or have been determined from M$_{\rm{J}}$ versus Spectral type relations from Dahn et al. (2002) for I-J$\geq$2.8 and Leggett (1992) for I-J$<$2.8.  The potential MG membership for each candidate is shown in the last column.}
\label{mgcandtable}
\begin{tabular}{lccccccccccc}
																				  
\hline																							  
Name & J-mag & J-H & J-K$_{\rm{s}}$ & I-J$^{a}$ & R$^{a}$ & PM$_{\rm{RA}}$ & PM$_{\rm{DEC}}$  & Total $\sigma$$_{\rm{PM}}$ & Distance & Sp.~T & MG$^{c}$  \\
&&&&&& (mas/yr) & (mas/yr) & (mas/yr) & (pc)$^{b}$ && Candidature\\														  
																								  
\hline 																							  
	
   SIPS 0004-5721              &    14.01$\pm$ 0.03    &    0.67$\pm$  0.04    &    1.07$\pm$    0.04    & 2.81$\pm$     0.30   &   19.78    &  157.0    &  -15.0    &  28.2    &   33 -- 53    &   M7.0  &  	    CA, HY, IC, PL     \\
   SIPS 0007-2458              &    13.11$\pm$ 0.02    &    0.67$\pm$  0.03    &    1.05$\pm$    0.03    & 2.66$\pm$     0.07   &   18.78    &  192.0    &  -59.0    &  13.6    &   23 -- 36    &   M7.0$^{g}$  &  	    CA, IC, PL         \\
   2MASS J00202315-2346053     &    12.35$\pm$ 0.02    &    0.63$\pm$  0.13    &    1.01$\pm$    0.03    & 2.31$\pm$     0.05   &   17.71    &  340.2    &  -65.0    &  20.2    &   17 -- 27    &   M6.0$^{g}$  &  	    IC		   \\
   DENIS-P J0021.0-4244        &    13.52$\pm$ 0.03    &    0.71$\pm$  0.04    &    1.22$\pm$    0.04    & 3.27$\pm$     0.11   &   19.46    &  262.4    &  -17.3    &  33.0    &   21 -- 34    & $\geq$M9.0$^{g}$  & 	    CA, HY, IC	   \\
   SIPS 0027-5401              &    12.36$\pm$ 0.02    &    0.65$\pm$  0.04    &    1.02$\pm$    0.03    & 2.46$\pm$     0.09   &   17.80    &  429.0    &  -18.0    &  33.0    &   17 -- 27    &   M7.0  &  	    HY, IC             \\
   SIPS 0030-6719              &    14.32$\pm$ 0.03    &    0.74$\pm$  0.05    &    1.20$\pm$    0.04    & 2.99$\pm$     0.16   &     ---    &  200.0    &    2.0    &  25.0    &   32 -- 50    &   M7.5  &  	    HY, IC, PL         \\
   SIPS 0031-3807              &    14.64$\pm$ 0.03    &    0.68$\pm$  0.05    &    1.05$\pm$    0.05    & 2.20$\pm$     0.10   &   19.17    &  117.0    &  -32.0    &  19.6    &   47 -- 74    &   M6.0  &  	    IC, PL             \\
   SIPS 0039-2256              &    14.30$\pm$ 0.03    &    0.67$\pm$  0.04    &    1.02$\pm$    0.05    & 2.41$\pm$     0.30   &   20.24    &  224.0    &   44.0    &  25.5    &   42 -- 66    &   M7.0  &  	    HY                 \\
   DENIS-P J004135.3-562112    &    11.96$\pm$ 0.02    &    0.64$\pm$  0.03    &    1.10$\pm$    0.03    & 2.48$\pm$     0.05   &   17.97    &   92.0    &  -62.1    &  21.0    &   12 -- 20    &   M7.5$^{g}$  & 	    CA, PL, SI	   \\
   SIPS 0054-4142              &    12.62$\pm$ 0.03    &    0.67$\pm$  0.04    &    1.00$\pm$    0.04    & 1.92$\pm$     0.05   &   17.11    &   64.0    & -108.0    &  17.5    &   20 -- 31    &   M5.0  &  	    CA                 \\
   SIPS 0058-0651              &    14.31$\pm$ 0.03    &    0.87$\pm$  0.04    &    1.41$\pm$    0.04    & 3.36$\pm$     0.05   &     ---    &  229.0    & -200.0    &  36.9    &   23 -- 36    &   L0$^{g}$  &    	    PL                 \\
   2MASS J01023244-2915370     &    15.63$\pm$ 0.06    &    0.66$\pm$  0.07    &    1.05$\pm$    0.10    & 2.14$\pm$     0.18   &   20.16    &  131.5    &   27.6    &  32.1    &   74 --118    &   M6.0  &  	    HY		   \\
   2MASS J01025334-3056300     &    14.70$\pm$ 0.04    &    0.62$\pm$  0.06    &    1.06$\pm$    0.06    & 2.30$\pm$     0.12   &   19.38    &  126.2    &   -1.6    &  21.6    &   47 -- 75    &   M6.5  &  	    HY, IC		   \\
   SIPS 0105-1826              &    13.26$\pm$ 0.02    &    0.68$\pm$  0.04    &    1.01$\pm$    0.03    & 1.92$\pm$     0.06   &   17.45    &  239.0    &   42.0    &  11.2    &   26 -- 42    &   M5.0  &  	    HY                 \\
   SIPS 0109-5100              &    12.23$\pm$ 0.02    &    0.69$\pm$  0.03    &    1.14$\pm$    0.03    & 3.04$\pm$     0.05   &   18.08    &  207.0    &   87.0    &  17.5    &   13 -- 21    &   M8.5$^{g}$  & 	    CA, HY, IC         \\
   NLTT 3868                   &    11.69$\pm$ 0.02    &    0.76$\pm$  0.03    &    1.27$\pm$    0.03    & 3.07$\pm$     0.30   &     ---    &  375.0    &   22.0    &  75.1    &     9.6$^{+0.2}_{-0.2}$$^{(d)}$ & M9.0$^{g}$  & 	      CA, HY, IC         \\
   LEHPM 1289                  &    13.60$\pm$ 0.03    &    0.65$\pm$  0.10    &    1.01$\pm$    0.03    & 2.40$\pm$     0.30   &   18.66    &  366.3    &   -4.4    &  35.4    &   31 -- 49    &   M7.0  & 	    HY		   \\
   SIPS 0115-2715              &    14.52$\pm$ 0.04    &    0.55$\pm$  0.04    &    1.00$\pm$    0.06    & 2.30$\pm$     0.31   &   18.74    &  149.0    &   32.0    &  12.2    &   48 -- 76    &   M6.5  & 	    HY                 \\
   2MASS J01230050-3610306     &    13.64$\pm$ 0.02    &    0.53$\pm$  0.06    &    1.45$\pm$    0.03    & 1.76$\pm$     0.08   &   17.58    &  124.5    &   57.4    &  30.5    &   15 -- 25    &   M4.5  & 	    HY, IC		   \\
   SIPS 0126-1946              &    14.26$\pm$ 0.03    &    0.65$\pm$  0.05    &    1.02$\pm$    0.05    & 2.01$\pm$     0.30   &   19.34    &  210.0    &  -11.0    &  47.0    &   41 -- 66    &   M5.5  & 	    HY                 \\
   LEHPM 1563                  &    12.66$\pm$ 0.03    &    0.62$\pm$  0.05    &    1.00$\pm$    0.03    & 3.13$\pm$     0.30   &   18.47    &  290.3    &  131.5    &  21.4    &   20 -- 32    &   M8.5  & 	    HY		   \\
   SIPS 0141-4232              &    14.67$\pm$ 0.04    &    0.68$\pm$  0.05    &    1.06$\pm$    0.05    & 2.35$\pm$     0.14   &     ---    &  108.0    &   19.0    &  20.4    &   47 -- 74    &   M6.5  & 	    HY, IC, PL         \\
   SIPS 0152-4003              &    15.26$\pm$ 0.05    &    0.77$\pm$  0.06    &    1.14$\pm$    0.09    & 2.55$\pm$     0.19   &     ---    &  192.0    &   26.0    &  26.3    &   54 -- 86    &   M7.0  & 	    HY                 \\
   2MASS J01531169-5122242     &    13.45$\pm$ 0.03    &    0.57$\pm$  0.05    &    1.03$\pm$    0.04    & 2.27$\pm$     0.08   &   18.41    &  123.2    &   41.0    &  21.4    &   28 -- 44    &   M6.5  & 	    HY, IC, PL	   \\
   2MASS J02041803-3945064     &    13.39$\pm$ 0.03    &    0.68$\pm$  0.04    &    1.10$\pm$    0.04    & 2.51$\pm$     0.13   &   18.81    &  119.1    &   16.6    &  23.5    &   24 -- 38    &   M7.0  & 	    CA, HY, IC, PL	   \\
   SIPS 0207-3721              &    12.44$\pm$ 0.03    &    0.61$\pm$  0.04    &    1.06$\pm$    0.04    & 2.63$\pm$     0.06   &   18.02    &  429.0    &  148.0    &  14.9    &     23.9$^{+8.4}_{-5.0}$$^{(e)}$    &  M7.0$^{g}$  &  HY                 \\
   SIPS 0212-6049              &    13.31$\pm$ 0.03    &    0.73$\pm$  0.04    &    1.15$\pm$    0.04    & 2.44$\pm$     0.09   &   18.61    &  112.0    &  -25.0    &  12.4    &   21 -- 34    &   M7.0  & 	    SI                 \\
   SIPS 0214-3237              &    14.01$\pm$ 0.03    &    0.67$\pm$  0.04    &    1.06$\pm$    0.04    & 2.22$\pm$     0.30   &     ---    &  148.0    &   42.0    &  34.2    &   34 -- 54    &   M6.0  & 	    HY, IC             \\
   2MASS J02155647-3456036     &    14.56$\pm$ 0.03    &    0.68$\pm$  0.04    &    1.07$\pm$    0.05    & 1.82$\pm$     0.09   &   18.65    &  160.9    &   55.9    &  34.6    &   44 -- 70    &   M5.0  & 	    HY		   \\
   SIPS 0219-7137              &    13.59$\pm$ 0.03    &    0.66$\pm$  0.04    &    1.08$\pm$    0.04    & 2.15$\pm$     0.09   &   18.64    &  173.0    &  109.0    &  21.1    &   27 -- 44    &   M6.0  & 	    HY                 \\
   SIPS 0223-4424              &    14.75$\pm$ 0.04    &    0.70$\pm$  0.06    &    1.02$\pm$    0.05    & 2.13$\pm$     0.30   &   19.32    &  156.0    &   80.0    &  15.7    &   51 -- 82    &   M6.0  & 	    HY                 \\
   2MASS J02262591-1839437     &    14.56$\pm$ 0.03    &    0.70$\pm$  0.06    &    1.12$\pm$    0.06    & 2.38$\pm$     0.31   &   20.23    &  193.2    &   67.0    &  36.6    &   40 -- 64    &   M6.5  & 	    HY		   \\
   SIPS 0235-0711              &    12.45$\pm$ 0.03    &    0.63$\pm$  0.04    &    1.02$\pm$    0.04    & 2.01$\pm$     0.06   &     ---    &  286.0    &   75.0    &  18.7    &   18 -- 28    &   M5.5$^{g}$  & 	    HY                 \\
   SIPS 0236-1616              &    14.43$\pm$ 0.03    &    0.69$\pm$  0.05    &    1.11$\pm$    0.04    & 2.66$\pm$     0.13   &     ---    &  130.0    &  -24.0    &  24.3    &   39 -- 61    &   M7.5  & 	    IC, PL, SI         \\
   SIPS 0239-1735              &    14.29$\pm$ 0.03    &    0.77$\pm$  0.05    &    1.25$\pm$    0.05    & 3.31$\pm$     0.16   &     ---    &   88.0    & -110.0    &  32.0    &   29 -- 46    &   L0$^{g}$  & 	    CA, PL, SI         \\
   2MASS J02495798-2147267     &    13.27$\pm$ 0.03    &    0.60$\pm$  0.03    &    1.02$\pm$    0.04    & 2.18$\pm$     0.30   &   18.59    &  150.9    &   31.2    &  37.4    &   26 -- 41    &   M6.0  & 	    HY, IC		   \\
   SIPS 0250-3056              &    15.47$\pm$ 0.07    &    0.69$\pm$  0.11    &    1.14$\pm$    0.11    & 2.20$\pm$     0.32   &   20.03    &  127.0    &   18.0    &  29.3    &   60 -- 95    &   M6.0  & 	    HY                 \\
   SIPS 0255-7711              &    14.96$\pm$ 0.04    &    0.68$\pm$  0.07    &    1.05$\pm$    0.08    & 1.90$\pm$     0.14   &   19.50    &   87.0    &  101.0    &  29.1    &   55 -- 87    &   M5.0  & 	    HY                 \\
   SIPS 0313-1347              &    14.41$\pm$ 0.02    &    0.71$\pm$  0.03    &    1.08$\pm$    0.05    & 1.83$\pm$     0.10   &   18.73    &  100.0    &   14.0    &  15.1    &   40 -- 64    &   M5.0  & 	    HY                 \\
   SIPS 0317-6415              &    14.81$\pm$ 0.04    &    0.64$\pm$  0.06    &    1.03$\pm$    0.05    & 1.86$\pm$     0.09   &   18.88    &   69.0    &   89.0    &  16.4    &   52 -- 83    &   M5.0  & 	    HY                 \\
   SIPS 0324-0050              &    14.77$\pm$ 0.04    &    0.61$\pm$  0.06    &    1.03$\pm$    0.07    & 2.52$\pm$     0.31   &     ---    &  107.0    &  -37.0    &  19.9    &   51 -- 81    &   M7.0  & 	    SI                 \\
   SIPS 0329-5747              &    13.98$\pm$ 0.03    &    0.61$\pm$  0.04    &    1.03$\pm$    0.04    & 2.76$\pm$     0.30   &   19.46    &  140.0    &  109.0    &  19.2    &   36 -- 57    &   M7.5  & 	    HY, IC             \\
   2MASS J03341065-2130343     &    11.91$\pm$ 0.02    &    0.61$\pm$  0.04    &    1.05$\pm$    0.03    & 1.92$\pm$     0.30   &   16.90    &  123.6    &    7.2    &  20.0    &   13 -- 21    &   M6.0$^{g}$  & 	    IC		   \\
   SIPS 0336-1437              &    15.41$\pm$ 0.03    &    0.65$\pm$  0.06    &    1.05$\pm$    0.08    & 2.49$\pm$     0.31   &     ---    &  111.0    &   49.0    &  25.1    &   66 --105    &   M7.0  &       HY                 \\

\hline																						  
\end{tabular}																					  
\end{footnotesize}																				  
\end{table}																					  																					  
\end{landscape}																					  
																						  
\begin{landscape}																				  
																						  
\setcounter{table}{1}																				  
\begin{table}
\begin{footnotesize}
\caption{}
\begin{tabular}{lcccccccccccc}
\hline         

	Name & J-mag & J-H & J-K$_{\rm{s}}$ & I-J$^{a}$ & R$^{a}$ & PM$_{\rm{RA}}$ & PM$_{\rm{DEC}}$  & Total $\sigma$$_{\rm{PM}}$ & Distance & Sp.~T$^{c}$ & MG$^{d}$  \\
&&&&&& (mas/yr) & (mas/yr) & (mas/yr) & (pc)$^{b}$ && Candidature\\																											  
\hline 

   LP 944-20                   &    10.73$\pm$ 0.02    &    0.71$\pm$  0.04    &    1.18$\pm$    0.03    & 3.43$\pm$0.04  &   17.15    &  355.0    &  273.8    &  27.5    &   5.0$\pm$0.1$^{(g)}$      &   M9.0$^{(f)}$  & 	   CA, IC		        \\
   SIPS 0401-4253              &    13.57$\pm$ 0.03    &    0.67$\pm$  0.03    &    1.07$\pm$    0.04    & 2.47$\pm$0.10  &   18.74    &   76.0    &  -71.0    &  17.7    &   27 -- 44     &   M7.0  & 	   SI                 	   \\
   2MASS J04075679-5535544     &    13.23$\pm$ 0.03    &    0.64$\pm$  0.04    &    1.02$\pm$    0.04    & 2.07$\pm$0.07  &   17.99    &   38.2    &  143.4    &  17.2    &   25 -- 41     &   M5.5  & 	   IC		      \\
   SIPS 0424-6243              &    14.02$\pm$ 0.03    &    0.65$\pm$  0.04    &    1.01$\pm$    0.05    & 2.06$\pm$0.09  &   18.49    &   10.0    &  123.0    &  20.1    &   37 -- 60     &   M5.5  & 	   IC                 	   \\
   2MASS J04291842-3123568     &    10.87$\pm$ 0.02    &    0.66$\pm$  0.03    &    1.10$\pm$    0.03    & 2.44$\pm$0.04  &   16.35    &  106.1    &   69.3    &  23.6    &    7 -- 12     &   M7.0/L1$^{g}$ & 	   CA, IC		      \\
   SIPS 0436-3429              &    14.73$\pm$ 0.03    &    0.62$\pm$  0.06    &    1.01$\pm$    0.06    & 2.41$\pm$0.14  &   20.09    &   81.0    &   92.0    &  29.1    &   52 -- 82     &   M7.0  & 	   HY                     \\
   SIPS 0441-6413              &    14.76$\pm$ 0.03    &    0.69$\pm$  0.05    &    1.04$\pm$    0.05    & 2.25$\pm$0.12  &   19.04    &   13.0    &  114.0    &  25.2    &   50 -- 80     &   M6.5  & 	   IC                     \\
   2MASS J04454336-5321345     &    12.85$\pm$ 0.02    &    0.60$\pm$  0.06    &    1.00$\pm$    0.03    & 2.73$\pm$0.07  &   18.47    &  343.8    &  485.8    &  12.9    &   22 -- 35     &   M7.5  & 	   HY                     \\
   SIPS 0501-7705              &    14.09$\pm$ 0.03    &    0.61$\pm$  0.05    &    1.08$\pm$    0.05    & 2.84$\pm$0.11  &   20.37    &   67.0    &   91.0    &  17.2    &   34 -- 55     &   M7.0  & 	   PL                 	   \\
   2MASS J05023867-3227500     &    12.44$\pm$ 0.03    &    0.63$\pm$  0.19    &    1.00$\pm$    0.04    & 2.11$\pm$0.30  &   17.67    &   62.6    & -171.9    &  20.8    &   18 -- 29     &   M6.5$^{g}$  &    SI		  	   \\
   2MASS J05280562-5919471     &    14.56$\pm$ 0.04    &    0.53$\pm$  0.21    &    1.02$\pm$    0.06    & 2.09$\pm$0.11  &   19.49    &   77.1    &  152.7    &  29.4    &   47 -- 75     &   M5.5  & 	   HY		  	   \\
   2MASS J05390417-4708060     &    13.26$\pm$ 0.03    &    0.65$\pm$  0.04    &    1.04$\pm$    0.04    & 2.56$\pm$0.07  &   18.59    &   30.8    &  350.0    &  16.5    &   25 -- 40     &   M7.0  & 	   HY		  	   \\
   2MASS J06003375-3314268     &    13.20$\pm$ 0.03    &    0.78$\pm$  0.18    &    1.19$\pm$    0.04    & 2.23$\pm$0.10  &   18.73    &  -28.3    &  153.7    &  29.3    &   19 -- 30     &   M7.5$^{g}$ & 	   HY		  	   \\
   2MASS J06361472-3241315     &    13.33$\pm$ 0.02    &    0.61$\pm$  0.03    &    1.04$\pm$    0.03    & 2.71$\pm$0.06  &   18.43    &  -20.3    &   97.9    &  23.1    &   26 -- 41     &   M7.5  & 	   HY		      \\
   SIPS 0648-6537              &    13.06$\pm$ 0.03    &    0.63$\pm$  0.04    &    1.00$\pm$    0.04    & 2.50$\pm$0.09  &   18.53    &    9.0    & -199.0    &  14.0    &   24 -- 38     &   M7.0  & 	   SI                     \\
   SIPS 0650-5003              &    14.34$\pm$ 0.03    &    0.66$\pm$  0.04    &    1.01$\pm$    0.05    & 2.52$\pm$0.10  &   19.73    &    4.0    &  228.0    &  17.0    &   43 -- 69     &   M7.0  & 	   HY                     \\
   SIPS 0745-5554              &    15.44$\pm$ 0.06    &    0.54$\pm$  0.10    &    1.10$\pm$    0.10    & 2.32$\pm$0.16  &   20.66    &   14.0    &  125.0    &  29.2    &   63 --100     &   M6.5  & 	   HY                     \\
   SIPS 0752-6621              &    15.16$\pm$ 0.05    &    0.73$\pm$  0.07    &    1.04$\pm$    0.08    & 2.56$\pm$0.21  &   20.39    &    8.0    &  114.0    &  26.1    &   61 -- 96     &   M7.0  & 	   HY                     \\
   SIPS 0913-0426              &    14.65$\pm$ 0.05    &    0.72$\pm$  0.07    &    1.08$\pm$    0.07    & 2.85$\pm$0.14  &     ---    &  131.0    &   22.0    &  25.3    &   45 -- 71     &   M7.0  & 	   CA, HY, IC             \\
   SIPS 1039-4110              &    11.97$\pm$ 0.02    &    0.65$\pm$  0.03    &    1.02$\pm$    0.03    & 2.35$\pm$0.30  &     ---    &   25.0    & -173.0    &   8.1    &   14 -- 22     &   M6.0 $^{g}$& 	   SI                     \\
   SIPS 1124-2019              &    13.44$\pm$ 0.03    &    0.62$\pm$  0.04    &    1.00$\pm$    0.04    & 2.72$\pm$0.07  &   19.17    &  129.0    &  -19.0    &  27.3    &   29 -- 46     &   M7.0  & 	   CA, IC                 \\
   SIPS 1228-1547              &    14.38$\pm$ 0.03    &    1.03$\pm$  0.04    &    1.61$\pm$    0.04    & 3.51$\pm$0.20  &     ---    &  250.0    & -331.0    &  64.2    &   17 -- 27     &  L1  & 	   PL                 	   \\
   DENIS-P J125052.6-212113    &    11.16$\pm$ 0.02    &    0.61$\pm$  0.04    &    1.03$\pm$    0.03    & 2.62$\pm$0.04  &   16.67    &  456.5    & -319.9    &  24.1    &    9 -- 15     &   M7.5$^{g}$ & 	   PL		      \\
   SIPS 1314-3212              &    13.76$\pm$ 0.03    &    0.62$\pm$  0.04    &    1.01$\pm$    0.04    & 2.84$\pm$0.30  &   19.54    &  142.0    & -160.0    &  28.3    &   33 -- 53     &   M7.0  & 	   PL                     \\
   SIPS 1329-4147              &    13.65$\pm$ 0.02    &    0.85$\pm$  0.04    &    1.38$\pm$    0.03    & 3.30$\pm$0.09  &   19.30    &  293.0    & -302.0    &  33.9    &   17 -- 28     &   M9.0$^{g}$  & 	   PL                     \\
   SIPS 1408-0435              &    15.00$\pm$ 0.04    &    0.81$\pm$  0.06    &    1.17$\pm$    0.07    & 2.34$\pm$0.20  &     ---    &   50.0    &  -97.0    &  23.7    &   45 -- 72     &   M6.5  & 	   PL                     \\
   SIPS 1450-1353              &    14.47$\pm$ 0.03    &    0.65$\pm$  0.06    &    1.02$\pm$    0.06    & 2.74$\pm$0.13  &   19.86    &   52.0    & -118.0    &  14.3    &   45 -- 72     &   M7.5  & 	   PL                 	   \\
   2MASS J15072779-2000431     &    11.71$\pm$ 0.02    &    0.67$\pm$  0.09    &    1.05$\pm$    0.03    & 2.72$\pm$0.07  &   17.33    &  114.2    &  -76.7    &  21.8    &   12 -- 19     &   M7.5$^{g}$ & 	   CA, IC, PL, SI	      \\
   SIPS 1632-0631              &    12.74$\pm$ 0.02    &    0.70$\pm$  0.03    &    1.12$\pm$    0.03    & 2.86$\pm$0.05  &     ---    &   29.0    & -366.0    &  37.1    &   17 -- 27     &   M7.0$^{g}$ & 	   PL                     \\
   SIPS 1758-6811              &    13.99$\pm$ 0.03    &    0.63$\pm$  0.04    &    1.00$\pm$    0.05    & 2.03$\pm$0.08  &   18.59    &    3.0    & -182.0    &  14.0    &   37 -- 59     &   M5.5  & 	   HY                     \\
   SIPS 1826-6542              &    10.57$\pm$ 0.03    &    0.61$\pm$  0.03    &    1.02$\pm$    0.03    & 2.34$\pm$0.30  &   16.22    &    6.0    & -311.0    &   9.0    &    7 -- 11     &   M6.5  & 	   CA, HY, IC, PL         \\
   SIPS 1903-3150              &    13.16$\pm$ 9.00    &    0.70$\pm$ 12.73    &    1.03$\pm$    0.04    & 1.68$\pm$0.06  &     ---    &   91.0    & -128.0    &  33.0    &   24 -- 38     &   M4.5  & 	   PL                     \\
   SIPS 1930-1943              &    12.34$\pm$ 0.03    &    0.65$\pm$  0.04    &    1.07$\pm$    0.04    & 2.48$\pm$0.05  &   17.85    &  242.0    &  -89.0    &  26.6    &   15 -- 25     &   M6.5$^{g}$ & 	   HY                     \\
   SIPS 1949-7136              &    13.92$\pm$ 0.03    &    0.80$\pm$  0.04    &    1.18$\pm$    0.04    & 2.69$\pm$0.09  &   18.43    &   36.0    & -183.0    &  26.5    &   27 -- 43     &   M7.5  & 	   HY, IC, PL             \\
   SIPS 1955-4137              &    14.86$\pm$ 0.04    &    0.75$\pm$  0.05    &    1.12$\pm$    0.06    & 3.09$\pm$0.21  &   20.38    &   73.0    &  -98.0    &  21.4    &   46 -- 73     &   M8.0  & 	   CA, HY, IC, PL         \\
   SIPS 2000-7523              &    12.73$\pm$ 0.03    &    0.77$\pm$  0.04    &    1.22$\pm$    0.04    & 2.95$\pm$0.07  &   18.05    &  179.0    &  -85.0    &  25.5    &   14 -- 23     &   M9.0$^{g}$ & 	   CA                 	   \\
   2MASS J20012463-5949000     &    13.24$\pm$ 0.02    &    0.64$\pm$  0.03    &    1.03$\pm$    0.04    & 2.26$\pm$0.06  &   18.61    &  163.3    &  -54.4    &  28.1    &   25 -- 40     &   M6.5  & 	   CA		      \\
   SIPS 2014-2016              &    12.54$\pm$ 0.02    &    0.69$\pm$  0.03    &    1.09$\pm$    0.04    & 2.78$\pm$0.30  &     ---    &  248.0    & -112.0    &  31.8    &   16 -- 26     &   M7.5$^{g}$ & 	   CA, HY, IC         	   \\
   2MASS J20312749-5041134     &    13.34$\pm$ 0.03    &    0.65$\pm$  0.09    &    1.03$\pm$    0.03    & 2.00$\pm$0.08  &   18.06    &  161.4    & -160.3    &  37.4    &   26 -- 42     &   M5.5  & 	   HY		      \\
   SIPS 2036-2537              &    14.21$\pm$ 0.03    &    0.62$\pm$  0.05    &    1.07$\pm$    0.05    & 2.56$\pm$0.09  &   19.91    &  131.0    & -134.0    &  15.6    &   37 -- 59     &   M7.0  & 	   PL                     \\
   SIPS 2039-1126              &    13.79$\pm$ 0.03    &    0.66$\pm$  0.05    &    1.11$\pm$    0.04    & 2.66$\pm$0.08  &   19.17    &   64.0    & -105.0    &  15.3    &   28 -- 45     &   M8.0$^{g}$ & 	   CA, PL                 \\
   SIPS 2041-1006              &    14.93$\pm$ 0.04    &    0.67$\pm$  0.05    &    1.08$\pm$    0.05    & 2.33$\pm$0.11  &     ---    &   57.0    &  -96.0    &  23.3    &   51 -- 81     &   M6.5  & 	   PL                     \\
   SIPS 2045-6332              &    12.62$\pm$ 0.03    &    0.81$\pm$  0.04    &    1.41$\pm$    0.04    & 3.33$\pm$0.10  &   19.03    &   97.0    & -201.0    &  18.8    &   10 -- 16     &   M9.5  &      CA, PL                 \\
   SIPS 2049-1716              &    11.81$\pm$ 0.03    &    0.60$\pm$  0.04    &    1.01$\pm$    0.03    & 2.19$\pm$0.30      &   16.93    &  369.0    & -142.0    &  44.0    &   13 -- 21     &   M7.0$^{g}$  &   HY, IC                 \\
   SIPS 2049-1944              &    12.85$\pm$ 0.02    &    0.63$\pm$  0.03    &    1.07$\pm$    0.03    & 2.59$\pm$0.05      &   18.58    &  179.0    & -279.0    &  13.0    &   20 -- 31     &   M7.5$^{g}$  &   PL                     \\
   SIPS 2049-3130              &    12.65$\pm$ 0.02    &    0.66$\pm$  0.03    &    1.06$\pm$    0.03    & 2.69$\pm$0.06      &   18.10    &  168.0    & -184.0    &  38.9    &   18 -- 29     &   M7.5 &   CA, IC, PL             \\
   SIPS 2057-1407              &    14.71$\pm$ 0.03    &    0.67$\pm$  0.05    &    1.05$\pm$    0.05    & 2.80$\pm$0.17      &     ---    &  179.0    &   12.0    &  30.1    &   48 -- 77     &   M7.5  &   HY                     \\

\hline
\end{tabular}
\end{footnotesize}
\end{table}

\end{landscape}

\begin{landscape}
\setcounter{table}{1}

\begin{table}
\begin{footnotesize}
\caption{}
\begin{tabular}{lccccccccccc}
        
\hline        
Name & J-mag & J-H & J-K$_{\rm{s}}$ & I-J$^{a}$ & R$^{a}$ & PM$_{\rm{RA}}$ & PM$_{\rm{DEC}}$  & Total $\sigma$$_{\rm{PM}}$ & Distance & Sp.~T$^{c}$ & MG$^{d}$  \\
&&&&&& (mas/yr) & (mas/yr) & (mas/yr) & (pc)$^{b}$ && Candidature\\																		  

\hline 

   SIPS 2100-6255              &    14.73$\pm$ 0.03    &    0.70$\pm$  0.04    &    1.05$\pm$    0.05    & 1.81$\pm$0.13      &     ---    &   65.0    & -110.0    &  19.7    &   49 -- 78     &   M5.0  &   HY                     \\
   SIPS 2105-1305              &    15.29$\pm$ 0.05    &    0.71$\pm$  0.08    &    1.24$\pm$    0.08    & 2.56$\pm$0.31      &     ---    &  114.0    &  -50.0    &  29.5    &   47 -- 74     &   M7.0  &   HY, IC                 \\
   2MASS J21062089-4044519     &    13.28$\pm$ 0.02    &    0.62$\pm$  0.09    &    1.01$\pm$    0.03    & 2.21$\pm$0.30      &   18.65    &  150.2    &  -83.1    &  35.9    &   27 -- 42     &   M6.0  &   HY, IC		      \\
   SIPS 2114-4339              &    13.02$\pm$ 0.02    &    0.66$\pm$  0.03    &    1.06$\pm$    0.03    & 2.55$\pm$0.09      &   18.61    &   49.0    & -148.0    &  18.0    &   21 -- 34     &   M7.0  &   CA                     \\
   SIPS 2118-1441              &    14.65$\pm$ 0.04    &    0.63$\pm$  0.06    &    1.03$\pm$    0.06    & 2.24$\pm$0.11      &   19.05    &  114.0    &  -66.0    &  17.5    &   48 -- 77     &   M6.0  &   PL                     \\
   SIPS 2119-0740              &    14.07$\pm$ 0.03    &    0.65$\pm$  0.05    &    1.05$\pm$    0.04    & 2.55$\pm$0.09      &     ---    &  152.0    & -117.0    &  19.2    &   35 -- 57     &   M7.0  &   IC, PL                 \\
   HB 2124-4228                &    13.32$\pm$ 0.02    &    0.66$\pm$  0.03    &    1.14$\pm$    0.03    & 2.76$\pm$0.09      &   19.35    &   97.2    & -170.4    &  49.2    &   35.7$\pm$10.1$^{f}$     &   M8.5$^{g}$ &   PL 	    \\
   SIPS 2128-3254              &    11.56$\pm$ 0.03    &    0.65$\pm$  0.05    &    1.05$\pm$    0.05    & 1.82$\pm$0.30      &   16.28    &  338.0    & -166.0    &  21.0    &   11 -- 18     &   M5.0  &   HY                     \\
   SIPS 2132-6248              &    14.45$\pm$ 0.03    &    0.69$\pm$  0.05    &    1.02$\pm$    0.05    & 2.30$\pm$0.12      &   19.08    &  100.0    &  -12.0    &  24.2    &   45 -- 71     &   M6.5  &   CA                     \\
   SIPS 2144-7518              &    14.53$\pm$ 0.04    &    0.62$\pm$  0.05    &    1.01$\pm$    0.06    & 2.23$\pm$0.13      &   20.11    &  102.0    & -124.0    &  19.8    &   47 -- 75     &   M6.0  &   HY                     \\
   SIPS 2151-4017              &    11.45$\pm$ 0.02    &    0.69$\pm$  0.03    &    1.03$\pm$    0.03    & 2.04$\pm$0.04      &   16.60    &  471.0    & -283.0    &  32.8    &   11 -- 17     &   M5.5  &   HY                     \\
   DENIS-P J220002.0-30383     &    13.44$\pm$ 0.03    &    0.79$\pm$  0.05    &    1.24$\pm$    0.04    & 2.76$\pm$0.30      &   19.50    &  198.7    &  -66.4    &  45.8    &   20 -- 31     &   M9.0/L0$^{g}$  &   CA, HY, IC	      \\
   SIPS 2200-2756              &    13.73$\pm$ 0.03    &    0.66$\pm$  0.04    &    1.03$\pm$    0.04    & 2.24$\pm$0.30      &   18.54    &  156.0    &   -8.0    &  16.0    &   31 -- 50     &   M6.0  &   HY                     \\
   2MASS J22071031-6917142     &    13.71$\pm$ 0.03    &    0.66$\pm$  0.05    &    1.02$\pm$    0.04    & 2.55$\pm$0.14      &   19.00    &  130.8    &  -50.7    &  29.9    &   31 -- 50     &   M7.0  &   CA, HY, IC, PL	      \\
   LEHPM 4480                  &    13.56$\pm$ 0.03    &    0.61$\pm$  0.04    &    1.06$\pm$    0.04    & 2.19$\pm$0.09      &   18.16    &  259.8    & -135.8    &  17.0    &   28 -- 44     &   M6.0  &   HY		      \\
   SIPS 2219-3944              &    15.38$\pm$ 0.05    &    0.70$\pm$  0.07    &    1.03$\pm$    0.08    & 2.01$\pm$0.16      &   19.47    &  106.0    &  -57.0    &  23.7    &   67 --107     &   M5.5  &   HY                     \\
   2MASS J22220368-4919234     &    14.55$\pm$ 0.03    &    0.60$\pm$  0.05    &    1.01$\pm$    0.05    & 2.10$\pm$0.30      &   19.66    &   66.8    & -122.4    &  32.5    &   48 -- 76     &   M6.0  &   PL		      \\
   SIPS 2227-4238              &    14.60$\pm$ 0.03    &    0.69$\pm$  0.05    &    1.02$\pm$    0.06    & 2.14$\pm$0.13      &   19.22    &  103.0    &  -59.0    &  22.0    &   48 -- 76     &   M6.0  &   HY, IC, PL             \\
   2MASS J22310865-4443184     &    13.00$\pm$ 0.02    &    0.64$\pm$  0.05    &    1.02$\pm$    0.03    & 2.07$\pm$0.30      &   17.77    &  163.2    &   -9.5    &  32.4    &   23 -- 36     &   M5.5  &   CA, HY, IC	      \\
   LEHPM 4908                  &    12.68$\pm$ 0.02    &    0.60$\pm$  0.03    &    1.04$\pm$    0.03    & 2.36$\pm$0.06      &   17.74    &  219.7    &  -63.1    &  14.4    &   19 -- 30     &   M6.5  &   HY, IC		      \\
   2MASS J22424129-2659272     &    13.35$\pm$ 0.03    &    0.66$\pm$  0.08    &    1.03$\pm$    0.04    & 2.17$\pm$0.09      &   17.91    &   99.3    &  -26.1    &  20.1    &   26 -- 42     &   M6.0  &   CA, HY, IC	      \\
   2MASS J22485685-6224060     &    14.91$\pm$ 0.05    &    0.72$\pm$  0.05    &    1.06$\pm$    0.08    & 2.00$\pm$0.31      &   19.97    &  128.7    & -102.2    &  40.3    &   52 -- 83     &   M5.5  &   HY		      \\
   2MASS J22545811-3228522     &    13.58$\pm$ 0.03    &    0.63$\pm$  0.04    &    1.01$\pm$    0.04    & 2.01$\pm$0.30      &   18.55    &   55.4    &  -83.9    &  23.4    &   30 -- 48     &   M5.5  &   CA, PL		      \\
   SIPS 2302-3939              &    15.39$\pm$ 0.05    &    0.69$\pm$  0.09    &    1.09$\pm$    0.08    & 2.07$\pm$0.16      &   19.98    &  118.0    &  -49.0    &  22.8    &   62 -- 99     &   M5.5  &   HY                     \\
   SIPS 2307-3910              &    15.08$\pm$ 0.05    &    0.78$\pm$  0.06    &    1.25$\pm$    0.06    & 2.75$\pm$0.19      &     ---    &  215.0    &   -6.0    &  21.0    &   42 -- 66     &   M7.5  &   HY                     \\
   2MASS J23113033-5256301     &    14.18$\pm$ 0.03    &    0.62$\pm$  0.05    &    1.01$\pm$    0.05    & 2.30$\pm$0.30      &   19.11    &  141.4    &  -43.8    &  29.3    &   40 -- 64     &   M6.5  &   HY, IC, PL	      \\
   SIPS 2318-4919              &    13.76$\pm$ 0.03    &    0.69$\pm$  0.04    &    1.07$\pm$    0.04    & 2.30$\pm$0.30      &   19.29    &  227.0    &  -25.0    &  22.1    &   30 -- 47     &   M8.0$^{g}$  &   HY                     \\
   2MASS J23214341-6106353     &    13.39$\pm$ 0.02    &    0.63$\pm$  0.10    &    1.00$\pm$    0.04    & 1.71$\pm$0.07      &   18.25    &  110.7    &  -58.7    &  16.9    &   28 -- 45     &   M4.5  &   IC, PL		      \\
   SIPS 2322-6357              &    14.26$\pm$ 0.03    &    0.61$\pm$  0.04    &    1.06$\pm$    0.05    & 2.19$\pm$0.13      &   19.39    &  122.0    &  -28.0    &  18.4    &   39 -- 61     &   M7.5$^{g}$  &   HY, IC, PL             \\
   SIPS 2326-3708              &    14.97$\pm$ 0.05    &    0.68$\pm$  0.07    &    1.03$\pm$    0.08    & 2.66$\pm$0.31      &     ---    &  158.0    &  -19.0    &  23.2    &   56 -- 89     &   M7.5 &   HY, IC                 \\
   SIPS 2339-5038              &    15.52$\pm$ 0.08    &    0.68$\pm$  0.10    &    1.05$\pm$    0.11    & 2.06$\pm$0.19      &   20.12    &  137.0    &   16.0    &  27.2    &   70 --112     &   M5.5  &   HY                     \\
   SIPS 2341-3550              &    13.53$\pm$ 0.03    &    0.55$\pm$  0.04    &    1.10$\pm$    0.04    & 2.55$\pm$0.08      &   18.86    &  154.0    &  -28.0    &  21.4    &   26 -- 41     &   M7.0  &   CA, HY, IC             \\
   SIPS 2343-2947              &    13.59$\pm$ 0.02    &    0.64$\pm$  0.03    &    1.08$\pm$    0.04    & 2.76$\pm$0.30      &   19.09    &  257.0    &   -8.0    &  22.0    &   27 -- 44     &   M7.5  &   HY, IC                 \\
   2MASS J23445797-6809398     &    13.98$\pm$ 0.03    &    0.62$\pm$  0.05    &    1.02$\pm$    0.04    & 2.50$\pm$0.10      &   19.43    &  177.6    &  -83.8    &  30.6    &   36 -- 58     &   M6.5$^{g}$  &   PL		      \\
   SIPS 2347-1821              &    13.07$\pm$ 0.03    &    0.60$\pm$  0.04    &    1.02$\pm$    0.04    & 2.23$\pm$0.30      &   18.18    &  219.0    &   41.0    &  17.3    &   23 -- 38     &   M6.0  &   HY                     \\
   SIPS 2348-3136              &    14.03$\pm$ 0.03    &    0.73$\pm$  0.04    &    1.16$\pm$    0.04    & 2.80$\pm$0.30      &   19.52    &  231.0    &  -54.0    &  27.7    &   30 -- 47     &   M7.5  &   IC                     \\
   SIPS 2350-6915              &    14.20$\pm$ 0.03    &    0.71$\pm$  0.05    &    1.07$\pm$    0.05    & 2.49$\pm$0.10      &   19.23    &  164.0    &  -11.0    &  26.1    &   37 -- 59     &   M7.0  &   HY, IC, PL             \\
   2MASS J23524913-2249295     &    13.02$\pm$ 0.02    &    0.70$\pm$  0.07    &    1.07$\pm$    0.04    & 2.64$\pm$0.30      &   18.33    &  222.0    & -175.0    &  32.5    &   21 -- 34     &   M7.5  &   PL		      \\
   SIPS 2353-4123              &    14.39$\pm$ 0.03    &    0.65$\pm$  0.05    &    1.06$\pm$    0.05    & 2.39$\pm$0.10      &   19.02    &  148.0    &   -1.0    &  23.0    &   41 -- 65     &   M6.0$^{g}$ &   HY, IC                 \\
   SIPS 2354-4210              &    15.30$\pm$ 0.05    &    0.62$\pm$  0.07    &    1.07$\pm$    0.08    & 1.89$\pm$0.31      &   19.48    &  111.0    &  -28.0    &  25.7    &   61 -- 97     &   M5.0  &   HY                     \\
   LEHPM 6542                  &    13.32$\pm$ 0.02    &    0.66$\pm$  0.15    &    1.11$\pm$    0.04    & 2.64$\pm$0.30      &   18.41    &  184.9    &   -4.6    &  31.4    &   23 -- 37     &   M7.5  &   CA, HY, IC             \\

\hline     																	 	  			  
\end{tabular}
\\																					  
CA=Castor, IC=IC2391, PL=Pleiades, HY=Hyades, SI=Sirius.\\
$^{a}$R-band uncertainties $\sim$~0.3~mag (Hambly, Irwin $\&$ MacGillivray 2001).\\	
$^{b}$Photometric distance from Section \ref{mgcandidates}.\\
$^{c}$Moving groups from Section \ref{mgcandidates}.\\
$^{d}$Costa et al. (2005). $^{e}$Costa et al. (2006). $^{f}$(Tinney 1996) $^{g}$SpT in literature - References:   Tinney (1998), Kirkpatrick et al. (2000), Burgasser et al. (2002), Cruz \& Reid (2002), Gizis et al. (2002), Cruz et al. (2003), Lodieu et al (2005), Siegler et al. (2005), Burgasser \& McElwain (2006), Basri \& Reiners (2006), Phan-Bao \& Bessel (2006),  \\																			  
\end{footnotesize}
\end{table}

\end{landscape}	

\section{Candidate Follow-Up}

\subsection{High Resolution Spectroscopy}

We observed seven objects using the FEROS echelle spectrometer mounted on the 2.2-m telescope based at La Silla (Chile) on the nights of 2006 December 29-30 from programme 078.C-0333(A) and 2007 June 16-17 from programme 079.C-0255(A).  FEROS covers the wavelength 3500-9200\AA\ over 39 orders with a resolving power of 48000 (Kaufer \& Pasquini 1998).  It is equipped with a double filter system allowing the object and the sky to be simultaneously measured.  We used exposure times that provided us with spectra of SNR$\sim$15-20 in the redder orders.  In both observing runs we experienced thin cloud and some time was lost to bad weather.  Seeing was generally 1-1.3 arcsecond.    

Starlink software was used for bias subtraction, flat-field division and optimal extraction of the spectra.  Bias subtraction was performed using the debias routine and the mean of the bias frames taken as part of the standard calibration for FEROS.  The FIGARO routine BCLEAN allowed us to remove any cosmic ray contamination whilst sacrificing a minimal amount of signal.  Lamp illuminated flat-fields were used to normalise the response of the detector.  The routine ECHOMOP was used to trace and clip the orders, model the flat-field balance and sky background, and to extract the spectra.  Finally, extracted spectra were wavelength calibrated using the ThAr illuminated images obtained after each observation.

\subsection{Spectral Types}

\begin{table*}

\begin{minipage}{180mm}
\caption{Spectral ratios for the FEROS objects with corresponding spectral types shown in parenthesis.  Adopted spectral types are assessing the full range of ratio information available (see text).  Literature spectral types are included where available.}
\label{indices}
\begin{tabular}{lcccccccc}
\hline

                Name      &         SpT (lit.)           &        TiO1+TiO2  &       VO1+VO2   &          VO        &     VO-a          &     TiO5           &   SpT(adopt.) \\

\hline

DENIS 004135.3-562112      &      M7.5$^{a}$              &      3.33 (M6.0)  &    ---          &     ---            &      ---          &     0.36 (M8.5)    &    M7.5 \\
2MASS J03341065-2130343    &      M6.0$^{b}$              &      2.78 (M5.0)  &    ---          &     ---            &      1.98 (M4.5)  &     0.34 (M4.5)    &    M4.5 \\     
LP 944-20                  &      M9.0$^{c}$              &      3.16 (M9.0)  &    3.04 (M8.5)  &     ---            &      2.28 (M7.5)  &     0.52 (M9.0)    &    M8.5 \\ 
2MASS J04291842-3123568    &      M7.5/L1 bin.$^{d}$    &      3.45 (M6.5)  &    3.08 (M8.0)  &     ---            &      ---          &     0.20 (M6.0)    &    M7.0 \\  
DENIS 125052.6-212113      &      M7.5$^{e}$              &      3.64 (M8.0)  &    3.06 (M8.5)  &     ---            &      2.16 (M6.5)  &     0.18 (M6.5)    &    M7.5 \\     
2MASS J15072779-2000431    &      M7.5$^{b}$              &      3.56 (M6.5)  &    2.67 (M7.0)  &     1.11 (M8.0)    &      2.37 (M8.5)  &     0.15 (M6.5)    &    M7.5 \\
SIPS 2049-1716             &      M6.0$^{f}$              &      3.06 (M5.5)  &    2.42 (M6.0)  &     1.08 (M8.0)    &      2.30 (M8.0)  &     0.20 (M6.0)    &    M6.5 \\
SIPS 2128-3254             &      ---                     &      3.13 (M6.0)  &    2.56 (M6.5)  &     1.06 (M7.0)    &      2.18 (M6.5)  &     0.24 (M5.5)    &    M6.5 \\    

\hline
\end{tabular}
$^{a}$Phan-Bao \& Bessell (2006) $^{b}$Cruz et al. (2003) $^{c}$Tinney (1998a) $^{d}$Siegler et al. (2005) $^{e}$Lodieu et al. (2005) $^{f}$Crifo et al. (2005)

\end{minipage}
\end{table*}

To obtain spectral types we used a variety of different spectral indices.  Our spectra often had a low SNR at certain wavelengths so indices were chosen for regions with good signal.  The VO index from Kirkpatrick, Henry \& Simons et al. (1995), CaH from Kirkpatrick, Henry \& McCarthy (1991), VO-a and TiO5 from Cruz \& Reid (2002) and PC3, TiO1+TiO2 and VO1+VO2 from Mart\'{i}n et al. (1999) were calculated, giving a number of spectral types from which we took the mean to give an adopted spectral type.  It should be noted that the TiO1+TiO2 and VO1+VO2 indices can infer two different spectral types and were thus only considered after we calculated the spectral type using other indices.  The PC3 Ratio covered low-signal regions of spectra so is omitted.  Table \ref{indices} shows the indices and spectral types for our FEROS objects with spectroscopic distances calculated using a spectral type-absolute J-magnitude relation from Cruz et al. (2003).

\subsection{Space Motions}

\label{spacemotions}

\begin{figure}
\includegraphics[angle=90,width=8.4cm]{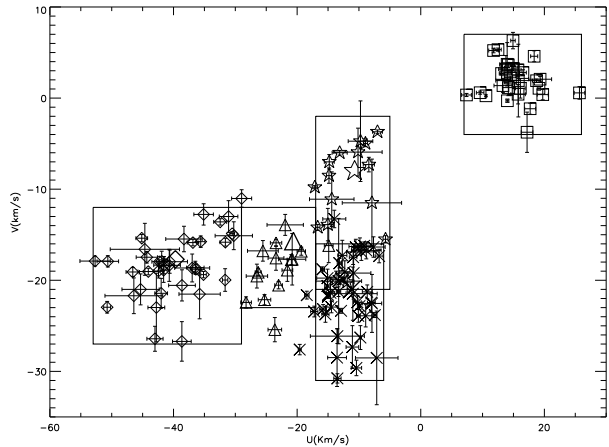}
\caption{$V$ versus $U$ space motions of samples of higher mass canonical moving group members from the literature (Leggett 1992; Dahn et al. 2002; Knapp et al. 2004; Vrba et al. 2004).  The moving group regions defined using the samples of higher mass members (see text) are indicated with boxes.  We plot members of the Castor, Sirius, Pleiades, IC2391 and Hyades moving groups as stars, squares, crosses, triangles and diamonds respectively. Representative moving group space motions (from Table 1) are also plotted for each group as larger symbols of the same type.}
\label{uvknown}
\end{figure}

\begin{figure}
\includegraphics[angle=90,width=8.4cm]{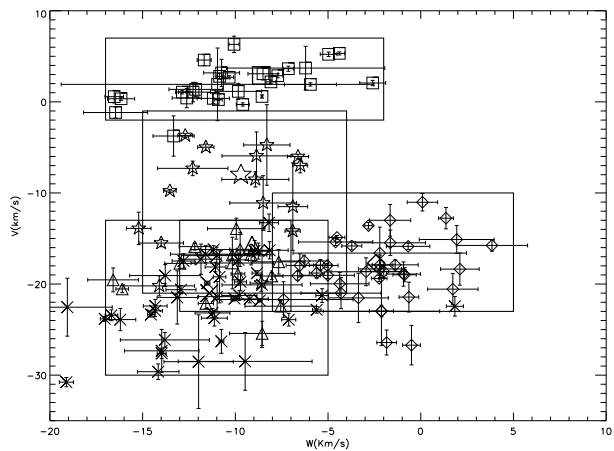}
\caption{The same as Figure \ref{uvknown}, except showing $V$ versus $W$ space motions.}
\label{wvknown}
\end{figure}

To obtain radial velocities we cross-correlated each individual order of the reduced spectra using an IDL routine that determines the digital cross-correlation of two arrays and finds the maximum cross-correlation function by fitting a Gaussian to the central peak.  This provided a set of radial velocities for each object from the different echelle orders with usable SNRs (5-7 orders per object).  Our final radial velocities and associated uncertainties came from the mean and standard deviation of these measurements.  A heliocentric correction was made to the measured shift between the object and the reference star so that the we could calculate radial velocities with respect to the Sun.

To allow for younger objects appearing intrinsically brighter, we created isochrones corresponding to the ages of our MGs and that of a field dwarf using atmospheric models from Barraffe et al (1998).  Approximating difference in absolute J-magnitude due to youth (according to each object's candidate group membership from Table \ref{mgcandtable}) was used in the Dahn distance-spectral type relation to give a new corrected distance assuming group membership.  Thus providing a number of distances for each of the radial velocity objects.  These distances (or any parallaxes available from the literature) were used in a case-by-case basis with proper motions and radial velocities to derive space motions for these objects using the transformation matrices of Johnson \& Soderblom (1987).  

To assess the potential MG membership of these objects, we compared their kinematics with the measured space motions of available samples of higher mass MG members from the literature (Montes et al. 2001; Barrado Y Navascu\'{e}s 1998).  Figures \ref{uvknown} and \ref{wvknown} show the $UV$ and $VW$ space motions of these higher mass samples. We trace boxes around each set of MG members (in each diagram), so as to contain a high fraction (with the occasional exception of one or two outliers) of the members. In this way we were able to effectively define the space motion ranges that are representative of each MG.   

\begin{figure}
\includegraphics[angle=90,width=8.4cm]{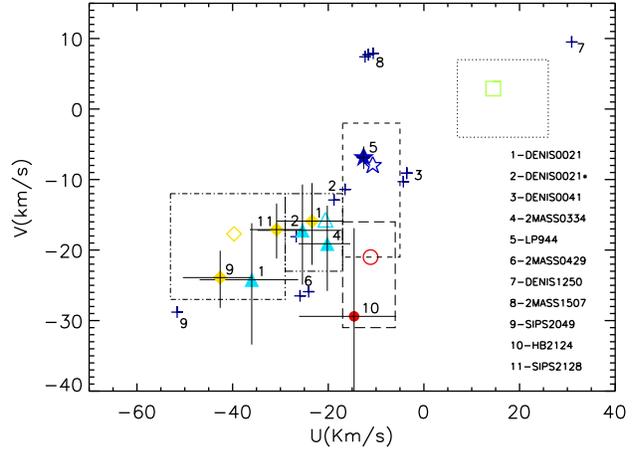}
\caption{ $V$ versus $U$ space motions of the 10 moving group candidates with radial velocities from our observations with FEROS and from the literature plotted as symbols corresponding to their kinematic membership.  Space motions that represent membership and non-membership are labelled with the respective object.  DENIS~0021 is plotted twice assuming it is a lone objects (object 1) and corrected for binarity (See section \ref{denispdata}; object 2).}
\label{radialuv}
\end{figure}

\begin{figure}
\includegraphics[angle=90,width=8.4cm]{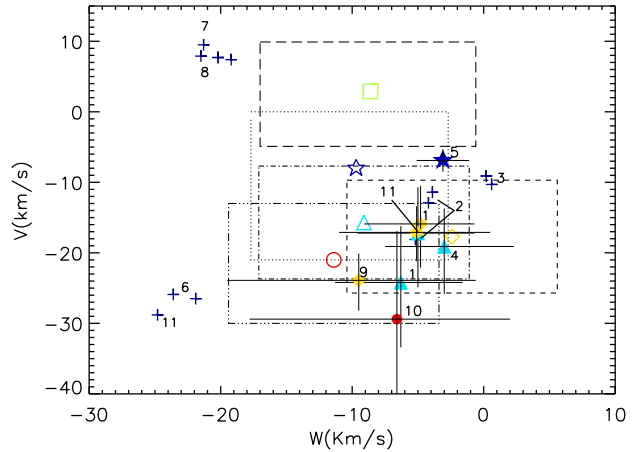}
\caption{The same as Figure \ref{radialuv}, except showing $V$ versus $W$.}
\label{radialwv}
\end{figure}

Table \ref{rvobjects} shows the radial velocity objects with their properties according to potential group membership in Table \ref{mgcandtable}.  Where, after evaluation, an object appears to be a kinematic member the values are highlighted in bold.  Figures \ref{radialuv} and \ref{radialwv} show the space motions of the 10 radial velocity objects (seven FEROS objects and three from the literature).  Kinematic members are plotted as symbols corresponding to their kinematic group membership, objects that are not kinematic members have all their possible space motions plotted as crosses.  We are able to reject four objects as possible MG members using our kinematic constraints, and thus have six candidates that are kinematically consistent with MG membership.  SIPS 2049-1716 and SIPS 2128-3254 are found to be kinematic members of the Hyades MG, 2MASS J03341065-2130343 of the IC2391 MG, LP 944-20 of the Castor MG and HB 2124-4228 of the Pleiades MG,  DENIS-P J0021.0-4244 is a possible member of the IC2391 or Hyades MGs.  We discuss our kinematic members below.

\begin{table*}

\begin{minipage}{160mm}
\caption{Radial velocities and derived $UVW$ space motions of 10 candidates from the MG candidate member list.  Rows shown in bold are adopted figures for that object due to kinematic results.}
\label{rvobjects}
\begin{tabular}{lcccccc}
\hline
Name & V$_{\rm{rad}}$ & distance  & U & V & W &  Moving \\
&   (kms$^{-1}$) & (pc)$^{f}$  & (kms$^{-1}$) & (kms$^{-1}$) & (kms$^{-1}$) & Group Candidate\\
\hline

DENIS-P J0021.0-4244     &2.0$^{a}$$\pm$3.0$^{b}$ &    \textbf{35.13$\pm$4.28}    &    \textbf{-36.0$^{9.7}_{10.9}$}  &   \textbf{-24.2$^{8.0}_{9.2}$}    &    \textbf{-6.3$^{4.7}_{5.0}$}  &  \textbf{IC2391}\\  
                         &                       &   26.170$\pm$3.19    &  -26.7$^{7.4}_{8.3}$    &  -18.1$^{6.1}_{7.0}$     &   -5.2$^{4.2}_{4.5}$   &   Castor\\           
                         &                       &    \textbf{23.00$\pm$2.80}    &    \textbf{-23.4$^{6.6}_{7.4}$}   &   \textbf{-15.9$^{5.4}_{6.2}$}    &    \textbf{-4.8$^{4.1}_{4.3}$}  & \textbf{Hyades}\\    
DENIS-P J0021.0-4244 (binary corr.)     &2.0$^{a}$$\pm$3.0$^{b}$ &   \textbf{24.90$\pm$4.28}      & \textbf{-25.4$^{8.2}_{9.4}$}      &   \textbf{-17.2$^{6.5}_{7.7}$}    &   \textbf{-5.0$^{4.3}_{4.6}$}   &   \textbf{IC2391}\\
                         &                       &   18.54$\pm$3.19     & -18.8$^{6.2}_{7.2}$      &  -12.9$^{4.9}_{5.9}$     &    -4.2$^{3.9}_{4.2}$  &    Castor\\
                         &                       &   16.30$\pm$2.80     & -16.5$^{5.6}_{6.4}$      &  -11.4$^{4.4}_{5.2}$     &    -3.9$^{3.8}_{4.0}$  &    Hyades\\
DENIS-P J004135.3-562112 &  2.4$\pm$1.0          &    20.80$\pm$2.32    &     -4.3$^{2.5}_{2.9}$   &   -10.3$^{3.1}_{3.6}$    &     0.6$^{2.1}_{1.9}$  & Pleiades\\           
                         &                       &    18.03$\pm$2.01    &     -3.6$^{2.2}_{2.6}$   &    -9.1$^{2.8}_{3.1}$    &     0.2$^{1.9}_{1.8}$  &   Castor\\           
                         &                       &    18.03$\pm$2.01    &     -3.6$^{2.2}_{2.6}$   &    -9.1$^{2.8}_{3.1}$    &     0.2$^{1.9}_{1.8}$  &   Sirius\\           
2MASS J03341065-2130343  &  19.0$\pm$0.8         &    \textbf{34.94$\pm$7.29}    &    \textbf{-20.2$^{4.8}_{6.0}$}   &   \textbf{-19.1$^{5.4}_{6.7}$}    &    \textbf{-3.0$^{5.3}_{4.5}$}  & \textbf{IC2391}\\    
LP 944-20                &  10.0$\pm$2.0$^{c}$   &     \textbf{5.04$\pm$0.11$^{d}$}    &    \textbf{-12.6$^{1.4}_{1.5}$}   &   \textbf{-6.9$^{1.6}_{1.6}$}     &   \textbf{-3.1$^{2.0}_{2.0}$}   & \textbf{Castor}\\    
2MASS J04291842-3123568  &  41.9$\pm$0.9         &    16.28$\pm$2.22    &    -25.9$^{2.6}_{3.0}$   &   -26.5$^{2.1}_{2.5}$    &   -21.9$^{2.7}_{2.4}$  &   IC2391\\           
                         &                       &    12.13$\pm$1.65    &    -24.1$^{2.0}_{2.4}$   &   -25.9$^{1.7}_{2.0}$    &   -23.6$^{2.2}_{1.9}$  &   Castor\\           
DENIS-P J125052.6-212113 &  -7.6$\pm$0.4         &    14.37$\pm$1.60    &     30.9$^{5.5}_{5.2}$   &     9.5$^{2.2}_{1.9}$    &   -21.3$^{2.8}_{3.0}$  & Pleiades\\           
2MASS J15072779-2000431  &   -22.2$\pm$1.3       &    18.53$\pm$2.06    &    -10.6$^{3.0}_{2.8}$   &     7.9$^{2.6}_{2.2}$    &   -21.5$^{3.2}_{3.6}$  & Pleiades\\           
                         &                       &    16.06$\pm$1.79    &    -11.6$^{2.7}_{2.5}$   &     7.7$^{2.3}_{2.0}$    &   -20.2$^{2.9}_{3.2}$  &   Castor\\           
                         &                       &    16.06$\pm$1.79    &    -11.6$^{2.7}_{2.5}$   &     7.7$^{2.3}_{2.0}$    &   -20.2$^{2.9}_{3.2}$  &   Sirius\\           
                         &                       &    14.12$\pm$1.57    &    -12.3$^{2.5}_{2.4}$   &     7.4$^{2.1}_{1.8}$    &   -19.2$^{2.6}_{2.9}$  &   Hyades\\           
SIPS 2049-1716           &   -35.4$\pm$2.4       &    28.54$\pm$4.81    &    -51.6$^{9.5}_{10.9}$  &   -28.8$^{5.3}_{6.1}$    &   -24.8$^{12.9}_{14.6}$ &   IC2391\\          
                         &                       &    \textbf{18.68$\pm$3.15}    &    \textbf{-42.6$^{6.8}_{7.8}$}   &   \textbf{-23.9$^{3.8}_{4.3}$}    &    \textbf{-9.5$^{8.9}_{10.0}$} & \textbf{Hyades}\\    
HB 2124-4228             &  -5.0$\pm$3.4$^{e}$   &    \textbf{35.70$\pm$10.10$^{e}$}   &   \textbf{-14.6$^{8.8}_{11.5}$}   &  \textbf{-29.4$^{12.5}_{15.8}$}    &    \textbf{-6.6$^{8.6}_{11.2}$} & \textbf{Pleiades}\\ 
SIPS 2128-3254           & -19.8$\pm$2.2         &    \textbf{16.61$\pm$2.80}    &    \textbf{-30.8$^{5.2}_{5.6}$}   &   \textbf{-17.1$^{3.7}_{4.1}$}    &    \textbf{-5.1$^{5.6}_{5.9}$}  & \textbf{Hyades}\\

\hline
\end{tabular}
$^{a}$Basri et al. (2000). $^{b}$Basri \& Reiners (2006). $^{c}$Tinney (1998a). $^{d}$Tinney (1996). $^{e}$ Tinney \& Reid (1998). $^{f}$ Spectroscopic distances unless indicated.
\end{minipage}
\end{table*}

\subsubsection{LP 944-20}

\label{lp944}

LP 944-20 (BRI 0337-3535) is a nearby (5pc; Dahn et al 2002) late M dwarf first identified by Luyten \& Kowal (1975). Spectroscopic observations (Tinney 1998) give a spectral type of M9, a $v_{\rm{rad}}$ of 10$\pm$2kms$^{-1}$, and place age constraints (using the lithium test; Magazzu, Mart\'{i}n, \& Rebolo 1993) of $\sim$475--650 Myr, suggesting a mass range $\sim$0.056--0.064M$_{\odot}$. It is well observed at different wavelengths (e.g. Cushing, Rayner \& Vacca 2005), with flaring X-ray emission detected (Rutledge et al. 2000), mid infrared emission measured (Apai et al. 2002), and both flaring and quiescent radio emission detected with the VLA (Berger et al. 2001). Recent spectroscopic analysis measures $v_{\rm{rot}}\sin{i}=30$kms$^{-1}$ (Zapatero Osorio et al. 2006), and shows a stable $v_{\rm{rad}}$ over a 6 day period, with no sign of close companions (Mart\'{i}n et al. 2006).

The measured parallax distance and $v_{\rm{rad}}$ allow accurate $UVW$ to be calculated, which are plotted in Figures \ref{radialuv} and \ref{radialwv}. Its position in these Figures clearly supports its membership of the Castor MG -- it lies very close to the canonical $UV$ space motion of this group, and is also consistent with this group's $VW$ space motion. Our analysis is entirely consistent with Ribas (2003), who previously identified LP944-20 as a likely member of the Castor MG. As a group member, the age of LP944-20 would be $\sim$320~Myr, which is reasonably consistent with the independently determined lithium age constraints. The metallicity could also be inferred as [M/H]=0.00$\pm$0.04 from Table 1, and the mass (from theoretical models for the assumed Castor age) as 0.063~M$_{\odot}$ (see Ribas 2003).  It should be noted that although these models act as a guide they are subject to significant uncertainties due to minimal consideration of dust opacity.  Temperatures can differ by $\sim$400K as shown by Figure 1 in Lawson \& Feigelson (2001), with models from Baraffe et al. (1998) and Siess, Dufour \& Forestini (2000) appearing comparatively cooler.  As a result of these known uncertainties note that our mass estimates (particularly those for the youngest candidates) come with the caveat that they could be subject to significant uncertainty, at the level of $\sim$30-80\% for ages from $\sim$600Myrs down to $\sim$50Myrs.

\subsubsection{DENIS-P J0021.0-4244}

\label{denispdata}

DENIS-P J0021.0-4244 (DENIS~0021 hereafter) was first spectral typed by Tinney \& Reid (1998), who estimated $\ge$M9 from low-resolution optical spectroscopy. Infrared HK spectroscopy was soon after presented by Delfosse et al. (1999). Basri et al. (2000) list it as an M9.5 dwarf, and use high resolution Cs~I and Rb~I spectra to estimate $T_{\rm eff}$ = 2300--2500K, $v_{\rm rot}\sin{i}$ = 17.5kms$^{-1}$ and $v_{\rm{rad}}$ = 2$\pm$1~kms$^{-1}$. Mohanty \& Basri (2003) also present a modest H$_{\rm{\alpha}}$ EW of 0.5\AA. Of particular interest is additional spectroscopy in Basri \& Reiners (2006) suggesting that DENIS~0021 could have radial velocity variations of 3.0$\pm$0.9kms$^{-1}$ (at the 2$\sigma$ level), and might thus be a spectroscopic binary.

Using the M$_{\rm{J}}$-spectral type relation from Dahn et~al. (2002) with correction for youth we obtained distance estimates of 23, 26 and 35~pc for the Hyades, Castor and IC2391 MGs respectively.  

Distance and radial velocity estimates show the object to be a possible kinematic member of the IC2391 or Hyades groups.  There is some ambiguity to this object's kinematic membership however, as the space motions that are calculated assuming a young object ($\sim$50~Myr) give a space motion that is more consistent with the relatively older Hyades MG ($\sim$600~Myr).  Similarly the space motions calculated assuming the object is a field dwarf place the object within the selection area of the younger IC2391 MG.  However, recalculating space motions assuming DENIS~0021 to be a spectroscopic binary (Object 2 on Figures \ref{radialuv} and \ref{radialwv}; see also Table \ref{rvobjects}), we find it has kinematic membership of IC2391 but no longer of the Hyades MG.  To determine the object's membership further follow up is required (see Section \ref{denislithiumtext}).

\begin{table*}
\begin{minipage}{173mm}
\caption{Approximate magnitude limits for different spectroscopic resolutions and wavebands are shown, with the potential applications of such spectroscopy for the specified SNR.  Magnitude faintness limits are created assuming 0.8 arcsecond seeing and a 2 hour exposure on an 8m class telescope.}
\label{exptimes}
\begin{tabular}{lccccc}
\hline
Resolution  & Waveband & Primary Applications  & SNR required & I limit & J limit \\ 
\hline

High          &  Optical    &  Radial velocities & 15 &  18.0 & -- \\
              &  Infrared   &  Radial velocities & 15 &  -- & 14.5 \\
\hline
Intermediate  &  Optical    &  Li detection, rotational velocities &  30 & 18.0 & -- \\
              &  Infrared   &  Rotational velocities  & 30 & -- & 14.5  \\
\hline
Low           &  Optical    &  Activity/age relation, g sensitivity, Li detection (L dwarfs only)  &  50 & 20.5 & -- \\
              &  Infrared   &  Activity/age relation, g sensitivity   &  50 & --  & 18.5 \\
\hline

\end{tabular}
\\
\end{minipage}
\end{table*}

\subsubsection{HB 2124-4228}

\label{hb2124}

Discovered by Hawkins \& Bessell (1988), HB 2124-4228 (hereafter HB~2124) is an M8.5 dwarf (Burgasser et al. 2002), with a measured parallax of 28$\pm$6mas/yr (Tinney 1996, see also Ianna 1993; van Altena Lee \& Hoffleit 1995) putting it at a distance of 36$^{+10}_{-7}$pc and having  $v_{\rm{rad}}$=-5.0$\pm$3.4kms$^{-1}$ (Tinney \& Reid 1998). The derived $UVW$ values are given in Table \ref{rvobjects} and plotted in Figures \ref{radialuv} and \ref{radialwv}. We note that our value for $V$ differs significantly from that presented by Tinney \& Reid (1998), who also derived space motions (albeit corrected slightly for solar motion) for this target. In addition, the $v_{\rm{tan}}$ listed in Tinney \& Reid (1998) is too high by comparison with the proper motion and the parallax, thus we will use the UVW values defined in this paper.

Photometric and astrometric selection placed this object as a candidate of the Pleiades MG and its space motions confirm kinematic membership.  If HB 2124 is a member of the Pleiades MG its age of 20--150~Myr would give a mass of $\sim$0.015-0.050~M$_{\odot}$.

\subsubsection{2MASS J03341065-2130343}

Cruz et al (2003) found this object to be an M6 dwarf.  We find this object to have a spectral type of M4.5 which shows some deviation from the Cruz value likely due to few usable indices for this object.  For this reason we use the Cruz spectral type when calculating space motions.  Correcting for youth to assume an IC2391 membership we estimate the object to have a distance of 34.9$\pm$7.3$\rm{pc}$ and space motions that confirm kinematic membership.  An IC2391 age would suggest a mass of $\sim$0.075~M$_{\odot}$.

\subsubsection{SIPS 2049-1716}

SIPS 2049-1716 is an unstudied object that photometric and astrometric tests suggest to be a Hyades candidate, thus inferring a distance estimate of 18.7$\pm$3.2$\rm{pc}$.  Galactic space motions confirm kinematic membership which would suggest a mass $>$0.080~M$_{\odot}$.

\subsubsection{SIPS 2128-3254}

SIPS 2128-3254 is also unstudied.  Galactic space motions consistent with a distance estimate of 16.6$\pm$2.8$\rm{pc}$ show the object to be a kinematic member of the Hyades, which, if confirmed we could infer a mass $>$0.080~M$_{\odot}$.

\subsection{Future Follow-Up and Additional Age Constraints}

\subsubsection{Age Determination}

To obtain radial velocities that are accurate to 1km/s it will be necessary to use high-resolution spectroscopy (R$\sim$30000).  This can be done optically in the I-band as well as the infrared as a high proportion of flux is emitted in these wavebands.   Table \ref{exptimes} shows the faintness magnitude limits for the planned spectroscopic follow-up.  For high-resolution optical and infrared spectroscopic follow-up the faintness magnitude limits encompass all our MG candidates. 

Although radial velocities can constrain our space motions enough to confirm kinematic members of our groups, determination of a candidate's age can finalise its group membership.  This can be achieved in such ways as:  Measurement of the Lithium 6708\AA\ doublet (e.g. Rebolo et al. 1996; Pavlenko et al. 2007); determining rotational velocities to differentiate between young and older M type ultracool dwarf populations (Reiners \& Basri. 2008); using gravity sensitive spectroscopic features to distinguish between young and old ultracool dwarfs (Gorlova et al. 2004; McGovern et al. 2004);  activity/age relation for late-type stars up to a spectral type of M7 (Mochnaki et al. 2002; Silvestri et al. 2006; Reiners and Basri 2008). 

Lithium absorption features are much less prominent in late M type objects than in L dwarfs so intermediate-resolution (R$\geq$8000) optical spectra will be required for the majority of our candidates (Manzi et al. 2008).  At this resolution we can also measure rotational velocities by measuring the broadening of absorption features. 

Our three L dwarfs will require only low-resolution spectra (R$\sim$500) with a good SNR to make lithium observations (Kirkpatrick et al. 1999).  This resolution can also provide gravity sensitive and activity/age relation constraints both for the late M and L dwarfs of our sample.

\begin{figure}
\includegraphics[angle=90,width=8.4cm]{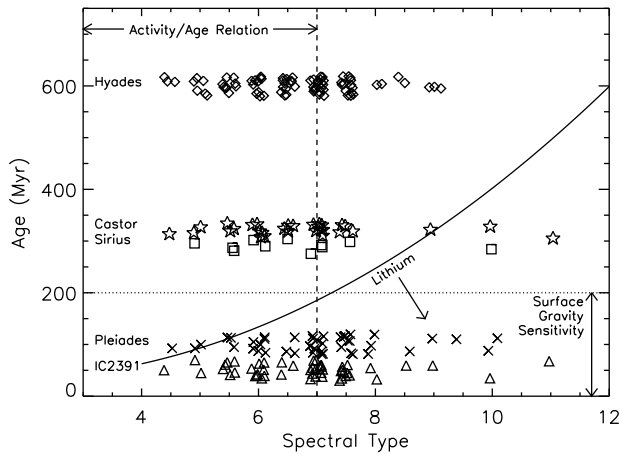}
\caption{Figure showing age versus spectral type for the candidates along with approximate follow-up limits.  The solid line marks the lithium edge with lithium depleted objects to the left.  The dashed line marks the limit of age/activity relation studies where objects to the left are plausible for testing.  The dotted line represents the upper age limit of testing for MG membership by gravity sensitive spectral features.  We plot members of the Castor, Sirius, Pleiades, IC2391 and Hyades moving groups as stars, squares, crosses, triangles and diamonds respectively.}
\label{lithium}
\end{figure}

Figure \ref{lithium} shows a plot of age (based upon potential group membership) versus spectral type (from Table \ref{mgcandtable}) for our candidates.  The figure illustrates which objects would be selected for the various methods of follow up mentioned previously.  Candidates that appear to the right of the lithium edge can be followed up with a lithium test programme, we have 51 objects that qualify for this.  There are also 51 candidates that appear younger than 200Myr so are eligible for follow up using spectroscopic gravity sensitive features.  We have 93 candidates left of the spectral type=~M7 limit and would thus be suitable for age/activity relation follow up, although candidates with a spectral type close to M7 may be subject to large uncertainties on their age.  Of our complete candidate list we have 23 objects that cannot be tested by any of these methods but will be eligible for age testing using rotational velocities.  Rotational velocities could also be used as an extra age constraint for the whole of our candidate list (see Jenkins et al. 2009).  Using spectroscopic methods we can apply at least one non-kinematic test to all our MG candidate providing a solid sample of MG members.

\subsubsection{Lithium Detections}

\label{denislithiumtext}

Lithium has previously been searched for and detected in LP944-20 (see Section \ref{lp944}).  In order to search for lithium in the ambiguous DENIS~0021 we obtained high resolution echelle spectra with UVES at the ESO VLT-UT2 Kueyen 8m telescope in July 2007.  We used the UVES standard setting centred on 760nm so that the 6707.8\AA\ lithium absorption line was covered.  In this region the resolution is R$\sim$40000 for a slit width of 1 arcsecond.  The spectra were extracted using the standard reduction procedures in the IRAF echelle package (bias subtraction, flat-field division and  optimal extraction of the spectra).  We obtained the wavelength calibration by taking spectra of a Th-Ar lamp. Finally, we corrected from telluric lines and normalised the spectra by a polynomial fit to the observed continuum.

\begin{figure}
\includegraphics[width=8.4cm]{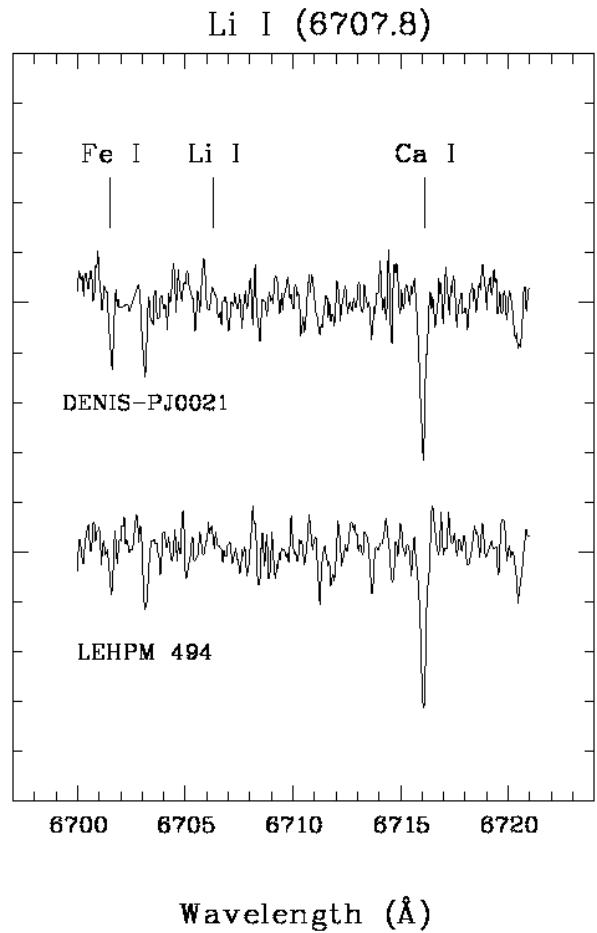}
\caption{ High resolution spectra obtained using UVES on the VLT for both components of the common proper motion pair DENIS~0021 and LEHPM 494 with a marker showing where the Li I 6707.8\AA\ absorption line is expected.  Other expected feature locations are overplotted.}
\label{lithiumnone}
\end{figure}

Figure \ref{lithiumnone} shows the high resolution spectra for the lithium region in DENIS~0021 and a common proper motion companion LEHPM 494 found at 78'' (previously discovered by Caballero (2007a;2007b)).  Visual inspection clearly reveals the lack of any visible absorption line.  Thus, we can say that the objects are either members of the Hyades MG or simply field dwarfs. Improved distance measurements for these objects would further constrain their galactic space motions.  Confirmation of Hyades MG membership would give an age of $\sim$600~Myr and masses of $\sim$0.060~M$_{\odot}$ and $\sim$0.095~M$_{\odot}$ for DENIS~0021 and LEHPM~494 respectively.

\section{Summary and Future Work}

We have created a sample of red low-mass objects with proper motions using the ELEHPM and SIPS catalogues.  From this we have found 132 MG candidates using photometric and astrometric techniques.  We assess the status of a subset of the catalogue using seven new radial velocities and three from the literature.  SIPS 2049-1716 and SIPS 2128-3254 are found to be kinematic members of the Hyades MG, 2MASS J03341065-2130343 of the IC2391 MG, LP 944-20 of the Castor MG and HB 2124-4228 of the Pleiades MG.  We find DENIS~0021 to be a possible kinematic member of the Hyades MGs after finding no evidence for lithium absorption within the object's spectra.

 The main tool for future work with our MG candidate sample will be the comprehensive measurement of radial velocities, to place accurate constraints on the space motions of the candidates, and identify bona-fide kinematic members of the five MGs under consideration. Additional spectroscopic analysis would also be desirable to place direct age constraints on kinematic members, such as the lithium test for the younger group candidates, the study of surface gravity sensitive spectral features, rotational velocities and chromospheric activity (e.g. G$\rm{\acute{a}}$lvez et al. 2006; 2007).  Such studies should produce a robust sample of up to $\sim$75 MG members (see Section \ref{mgcandidates}), representing a major new sample of young nearby ultracool dwarfs with well constrained ages and compositions.

This sample could form the basis for studies of low-gravity ultracool atmospheres and low-mass binary systems, as well as providing ideal targets for AO imaging searches for giant extra-solar planets, and further prospects of ultracool dwarf multiple planet systems via future near infrared radial velocity surveys (e.g. Jones et al. 2008).

\end{document}